\documentclass[%
 reprint,
 amsmath,amssymb,
 aps,
 prd,nofootinbib,
 superscriptaddress
]{revtex4-2}

\usepackage{silence}
\usepackage{xspace}
\usepackage{bm}

\usepackage{amsmath}
\usepackage{amssymb}
\usepackage{graphicx}
\usepackage{mathtools}
\usepackage[hidelinks]{hyperref}
\usepackage{subcaption}
\DeclareCaptionJustification{apsjustified}{\leftskip=0pt\rightskip=0pt\parfillskip=0pt plus 1fil}
\usepackage{cleveref}

\usepackage{xcolor}
\usepackage{placeins}
\usepackage{siunitx}

\begin{document}

\crefname{figure}{Fig.}{Figs.}
\Crefname{figure}{Figure}{Figures}
\crefname{equation}{Eq.}{Eqs.}
\Crefname{equation}{Equation}{Equations}
\crefname{section}{Sec.}{Secs.}
\Crefname{section}{Section}{Sections}
\crefname{table}{Tab.}{Tabs.}
\Crefname{table}{Table}{Tables}
\crefname{appendix}{App.}{Apps.}
\Crefname{appendix}{Appendix}{Appendices}
\creflabelformat{equation}{#2\textup{#1}#3}
\renewcommand{\crefrangeconjunction}{--}

\makeatletter

\newcommand{\coll}[2]{%
	$\coll@process{#1} + \coll@process{#2}$%
}

\newcommand{\collThree}[3]{%
	$\coll@process{#1} / \coll@process{#2} + \coll@process{#3}$%
}

\newcommand{\collFour}[4]{%
  $\coll@process{#1} / \coll@process{#2} / \coll@process{#3} + \coll@process{#4}$%
}

\newcommand{\coll@process}[1]{%
  \ifx#1A%
    #1%
  \else\ifx#1B%
    #1%
  \else\ifx#1x%
    #1%
  \else\ifx#1p%
    #1%
  \else\ifx#1d%
    #1%
  \else
    \coll@checkHeThree{#1}%
  \fi\fi\fi\fi\fi%
}

\newcommand{\coll@checkHeThree}[1]{%
  \ifnum\pdfstrcmp{#1}{He3}=0 %
    {}^3\mathrm{He}%
  \else
    \mathrm{#1}%
  \fi
}

\newcommand{\rt}{\ensuremath{\mathbf{r}_{\perp}}\xspace}
\newcommand{\bt}{\ensuremath{\mathbf{b}_{\perp}}\xspace}

\newcommand{\aaa}{\ensuremath{A + A}\xspace}

\newcommand{\pp}{\ensuremath{p + p}\xspace}
\newcommand{\ab}{\ensuremath{A + B}\xspace}
\newcommand{\ep}{\ensuremath{e + p}\xspace}
\newcommand{\pbpb}{\ensuremath{\mathrm{Pb} + \mathrm{Pb}}\xspace}
\newcommand{\pb}{\ensuremath{\mathrm{Pb}}\xspace}
\newcommand{\auau}{\ensuremath{\mathrm{Au} + \mathrm{Au}}\xspace}
\newcommand{\au}{\ensuremath{\mathrm{Au}}\xspace}
\newcommand{\oo}{\ensuremath{\mathrm{O} + \mathrm{O}}\xspace}
\newcommand{\ox}{\ensuremath{\mathrm{O}}\xspace}
\newcommand{\nene}{\ensuremath{\mathrm{Ne} + \mathrm{Ne}}\xspace}
\newcommand{\neon}{\ensuremath{\mathrm{Ne}}\xspace}
\newcommand{\arar}{\ensuremath{\mathrm{Ar} + \mathrm{Ar}}\xspace}
\newcommand{\ar}{\ensuremath{\mathrm{Ar}}\xspace}
\newcommand{\xexe}{\ensuremath{\mathrm{Xe} + \mathrm{Xe}}\xspace}
\newcommand{\xe}{\ensuremath{\mathrm{Xe}}\xspace}
\newcommand{\li}{\ensuremath{{}^6\mathrm{Li}}\xspace}
\newcommand{\lili}{\ensuremath{{}^6\mathrm{Li} + {}^6\mathrm{Li}}\xspace}

\newcommand{\pa}{\ensuremath{p + A}\xspace}
\newcommand{\da}{\ensuremath{d + A}\xspace}
\newcommand{\pdha}{\ensuremath{p / d / {}^3 \mathrm{He} + A}\xspace}
\newcommand{\pdhau}{\ensuremath{p / d / {}^3 \mathrm{He} + \mathrm{Au}}\xspace}

\newcommand{\ppb}{\ensuremath{p + \mathrm{Pb}}\xspace}
\newcommand{\po}{\ensuremath{p + \mathrm{O}}\xspace}
\newcommand{\pau}{\ensuremath{p + \mathrm{Au}}\xspace}
\newcommand{\dau}{\ensuremath{d + \mathrm{Au}}\xspace}
\newcommand{\dpb}{\ensuremath{d + \mathrm{Pb}}\xspace}

\newcommand{\hea}[1]{\ensuremath{{}^{#1}\mathrm{He} + A}\xspace}

\newcommand{\heau}[1]{\ensuremath{{}^{#1}\mathrm{He} + \mathrm{Au}}\xspace}

\newcommand{\hehe}[1]{\ensuremath{{}^{#1}\mathrm{He} + {}^{#1}\mathrm{He}}\xspace}

\newcommand{\he}[1]{\ensuremath{{}^{#1}\mathrm{He}}\xspace}

\newcommand{\sipg}{\ensuremath{\sigma_{\text{inel}}^{\text{IPG}}}\xspace}

\newcommand{\avg}[1]{\ensuremath{\left\langle #1 \right\rangle}\xspace}
\newcommand{\Ncoll}{\ensuremath{N_{\text{coll}}}\xspace}
\newcommand{\Nch}{\ensuremath{N_{\text{ch}}}\xspace}
\newcommand{\Sperp}{\ensuremath{S_{\perp}}\xspace}
\newcommand{\Qsminsq}{\ensuremath{Q_{s,\text{min}}^{2}}\xspace}

\makeatother

\title{Centrality-dependent nuclear modification from hard-soft correlations in the glasma}

\author{Coleridge Faraday}
\email{frdcol002@myuct.ac.za}
\affiliation{Department of Physics\char`,{} University of Cape Town\char`,{} Private Bag X3\char`,{} Rondebosch 7701\char`,{} South Africa}

\author{W.\ A.\ Horowitz}
\email{wa.horowitz@uct.ac.za}
\affiliation{Department of Physics\char`,{} University of Cape Town\char`,{} Private Bag X3\char`,{} Rondebosch 7701\char`,{} South Africa}
\affiliation{Department of Physics\char`,{} New Mexico State University\char`,{} Las Cruces\char`,{} New Mexico\char`,{} 88003\char`,{} USA}
\affiliation{Theoretical Sciences Visiting Program\char`,{} Okinawa Institute of Science and Technology Graduate University\char`,{} Onna\char`,{} 904-0495\char`,{} Japan}

\author{Bj\"orn Schenke}
\email{bschenke@bnl.gov}
\affiliation{Physics Department\char`,{} Brookhaven National Laboratory\char`,{} Upton\char`,{} NY 11973\char`,{} USA}

\date{\today}

\begin{abstract}
We present first predictions 
from the saturation-physics-based framework IP-Glasma
for the nuclear modification factor $R_{AB}$ as a function of centrality in \po, \ppb, \oo, and \pbpb collisions
due only to initial-state effects.
The same framework is responsible for both
soft ($p_T \lesssim 3 ~\mathrm{GeV}$) and semi-hard ($5 ~\mathrm{GeV} \lesssim p_T \lesssim 20 ~\mathrm{GeV}$) particle production,
enabling the study of initial-state correlations 
between bulk and intermediate-$p_T$ particle production from a first-principles framework.
We show that,
tuned only to HERA data,
IP-Glasma accurately predicts
the self-normalized multiplicity distributions in \po, \ppb, \oo, and \pbpb collisions;
the minimum-bias nuclear modification factor in \po and \ppb collisions;
the centrality-cut nuclear modification factor in \ppb collisions;
and the anomalous suppression of $R_{AA}$ observed in very peripheral \pbpb collisions.
We find that IP-Glasma predicts significantly less suppression than is measured
in both $\sqrt{s_{NN}} = 5.36 ~\mathrm{TeV}$ \oo and
$\sqrt{s_{NN}} = 5.02 ~\mathrm{TeV}$ \pbpb collisions at the Large Hadron Collider,
in qualitative agreement with the scenario in which $R_{AA} < 1$
is due to final-state energy loss. %
We show that the inelastic nucleon-nucleon cross section ($\sigma_{\text{inel}}^{NN}$) produced by IP-Glasma is extremely sensitive to the area of the subnucleonic hotspots;
the same values of the hotspot area that reproduce the measured $\sigma_{\text{inel}}^{NN}$ also reproduce minimum-bias $R_{pA}$.
Finally, 
we show that the hard-soft correlations in IP-Glasma
arise from event-by-event fluctuations in the color fields,
which simultaneously drive enhanced production of both soft and hard particles.

\end{abstract}

\maketitle

\section{Introduction}

There is overwhelming experimental evidence 
that a quark-gluon plasma (QGP) is formed in the ultrarelativistic collisions of heavy ions
at the Relativistic Heavy-Ion Collider (RHIC) 
and the Large Hadron Collider (LHC) \cite{PHENIX:2004vcz, Busza:2018rrf, ALICE:2022wpn}.
This evidence comes from the observation of QGP signatures including, for example,
the large, nonzero $v_2$ at low transverse momentum ($p_T$) associated with \emph{elliptic flow} \cite{STAR:2000ekf,PHENIX:2003qra,ALICE:2010suc},
the enhancement of particles containing strange quarks \cite{STAR:2003jis,ALICE:2013xmt},
and the suppression of high-$p_T$ jets, \emph{jet quenching} \cite{PHENIX:2001hpc,STAR:2003fka,ALICE:2010yje,ALICE:2012ab,CMS:2012aa}. 
More recently,
many of these same signatures have been observed in significantly smaller collision systems, such as \ppb collisions at the LHC \cite{ATLAS:2012cix, ALICE:2012eyl, CMS:2012qk, ALICE:2016sdt, ATLAS:2013jmi, ALICE:2014dwt,CMS:2015yux,CMS:2018loe,ATLAS:2019pvn,ALICE:2024vzv}
and \pdhau collisions at RHIC \cite{PHENIX:2018lia, STAR:2022pfn},
with the notable exception of jet quenching \cite{ALICE:2014xsp, ATLAS:2022iyq,CMS:2025jbv},
the so-called ``small-system energy loss puzzle.''

The standard candle for jet quenching is 
the nuclear modification factor $R^h_{AB}(p_T) \equiv (\Ncoll)^{-1} (d N_{AB}^{h} / d p_T)/(d N_{pp}^{h} / d p_T )$
for the collision \ab,
where $\Ncoll$ is the number of inelastic binary collisions,
typically calculated with the Glauber model \cite{Miller:2007ri},
and $d N_{AB}^h / d p_T$ $(d N_{pp}^h / d p_T)$ is the $p_T$-differential yield of hadrons $h$ in \ab (\pp) collisions.
In the absence of nuclear effects, $R^h_{AB} = 1$ by construction;
the measured $R_{AA} \ll 1$ in \auau and \pbpb collisions 
is interpreted as evidence for significant final-state partonic energy loss \cite{Wiedemann:2009sh,Cao:2020wlm}.

One of the difficulties with small systems is that most QGP signatures
are observed only in high-multiplicity events~\cite{Grosse-Oetringhaus:2024bwr},
in which high-$p_T$ measurements are sensitive to correlations
between hard- and soft-particle production.
In small systems, 
such hard-soft correlations are sourced by fluctuations at fixed impact parameter---%
including color fluctuations~\cite{Alvioli:2013vk,Alvioli:2014eda,Alvioli:2017wou,Perepelitsa:2024eik},
multi-parton interactions~\cite{Loizides:2017sqq},
subnucleonic geometry~\cite{Schenke:2020mbo},
saturation scale fluctuations~\cite{Mantysaari:2016ykx,Mantysaari:2022ffw},
and momentum conservation~\cite{Kordell:2016njg,JETSCAPE:2024dgu}---%
which also limit the applicability of the Glauber model
and the reliability of the $R_{pA}$ normalization \cite{ALICE:2014xsp,ATLAS:2016xpn}.
Whether these hard-soft correlations are purely selection biases 
that lead to more or fewer jets in the sample,
or whether the correlations are indicative of different types of plasmas that are formed when jets are present 
is an open question.
In practice,
hard-soft correlations and selection biases 
drive $R_{pA}$ significantly above unity in central
collisions and below unity in peripheral collisions~\cite{ALICE:2014xsp,ATLAS:2016xpn},
contrary to the scaling expected from energy loss \cite{Faraday:2024qtl,Faraday:2025pto,Faraday:2025prr}.
Together, these effects make both the experimental measurement and the theoretical prediction of jet quenching in small systems extremely difficult.

In 2025, 
the LHC performed a short run of \oo and \nene collisions with one of the main physics goals being to gain insight into the small-system energy loss puzzle
by providing a well-controlled small system where both elliptic flow and jet quenching might be observed \cite{Citron:2018lsq, Huss:2020dwe,Huss:2020whe, Brewer:2021kiv}.
Additionally, 
due to the symmetric nature of these collisions,
one anticipates less sensitivity to the structure of the proton
and any potential hard-soft correlations \cite{Faraday:2025prr}.
Measurements of these collisions successfully found evidence for both
elliptic flow \cite{ALICE:2025luc,ATLAS:2025nnt},
driven by the initial nuclear geometry in good agreement with hydrodynamics models \cite{Nijs:2021clz,Giacalone:2024luz,Mantysaari:2025tcg},
and jet quenching \cite{ALICE:2026zck,CMS:2025bta,CMS:2025ojt}, in good agreement with final-state partonic energy loss models \cite{Faraday:2025pto,Faraday:2025prr,Huss:2020dwe,Huss:2020whe,vanderSchee:2025hoe,Pablos:2025cli,Zakharov:2025mbk,Kudinoor:2026wcs},
providing strong evidence for quark-gluon plasma formation in \oo and \nene collisions.
As a consequence of pathlength-dependent energy loss, a nonzero high-$p_T$ $v_2$ is naively expected,
and measurements indeed find a significant signal~\cite{CMS:2026qth,ATLAS:2026mdt}.
Quantitative calculations, however,
find that energy loss cannot generate a $v_2$ of the measured size
in small systems including \oo collisions~\cite{Bert:2026uxa},
leaving the origin of the high-$p_T$ $v_2$ an open question.

Despite this progress, the small-system energy loss puzzle persists in asymmetric small systems: while a 
multitude of low-$p_T$ measurements indicative of QGP formation exist in 
high-multiplicity \pdha collisions, the high-$p_T$ picture remains ambiguous \cite{Grosse-Oetringhaus:2024bwr}.
Different experimental strategies designed to suppress the hard-soft 
correlations and centrality bias in $R_{pA}$
yield contradictory conclusions.
By using a centrality definition related to the number of spectator neutrons---%
which should be less sensitive to hard-soft correlations---%
ALICE measures $R_{pA}$ consistent with unity in central collisions \cite{ALICE:2014xsp},
inconsistent with predictions from energy loss models \cite{Faraday:2025pto},
even once corrections relevant in small systems are included \cite{Faraday:2023mmx,Kolbe:2015rvk}.
Similarly,
measurements of dijets by ATLAS \cite{ATLAS:2022iyq} and CMS \cite{CMS:2025jbv} using the same centrality definition 
find no additional back-to-back asymmetry that would indicate energy loss.
The PHENIX collaboration considers the self-normalized double ratio between the pion and direct-photon nuclear modification factors,
$R_{AA}^{\pi} / R_{AA}^{\gamma}$,
which is less sensitive to centrality bias
and Glauber model uncertainties \cite{PHENIX:2023dxl}.
Using this approach, PHENIX finds nontrivial suppression in central collisions,
consistent with theoretical predictions of energy loss~\cite{Faraday:2024qtl,Faraday:2025pto}.
Measurements of the high-$p_T$ $v_2$ in \ppb collisions find values larger than zero~\cite{ATLAS:2019vcm,CMS:2025kzg},
suggestive of energy-loss effects,
yet quantitative calculations of energy loss do not produce a large enough $v_2$ to describe these data~\cite{Bert:2026uxa}.
The apparent tension between these measurements 
may reflect the fact that each observable is differently sensitive to hard-soft correlations, 
and no existing theoretical framework can simultaneously account for 
soft particle production,
centrality selection,
and hard-probe suppression 
in a single consistent calculation in small systems.

Several previous approaches have attempted to model both hard and soft particle production,
but each captures only a subset of the appropriate physics in the $5 ~\mathrm{GeV} \lesssim p_T \lesssim 20 ~\mathrm{GeV}$ kinematic range relevant for small-system energy loss. 
Frameworks that correlate hard and soft modes through energy-momentum conservation~\cite{Kordell:2016njg,JETSCAPE:2024dgu}
invoke a mechanism that is not expected to dominate
for $5~\mathrm{GeV} \lesssim p_T \lesssim 20~\mathrm{GeV}$ particles at LHC energies,
where the jet $p_T$ constitutes a negligible fraction of the total energy available in the collision.
Indeed, 
energy-momentum conservation leads to a negative correlation between hard and soft particle production,
opposite to that observed in \ppb collisions in the aforementioned kinematic range \cite{ALICE:2014xsp}.
Models of hard-soft correlations driven by color fluctuations~\cite{Alvioli:2013vk,Alvioli:2014eda,Alvioli:2017wou}
require fairly significant parameter fitting
and are applicable only at large Bjorken $x$---outside the kinematic regime
currently relevant for partonic energy loss in small systems at LHC energies.
Modeling \ppb collisions as an incoherent superposition of $\Ncoll$ \pp collisions with event generators~\cite{ALICE:2014xsp,Loizides:2017sqq}
reproduces the centrality dependence of $R_{pA}$,
but predicts $\Ncoll$ scaling of soft particle production,
contrary to the observed approximate $N_{\text{part}}$ scaling~\cite{ALICE:2014xsp}.
Moreover, such incoherent-superposition-based implementations
do not provide a natural initial condition for hydrodynamic evolution,
which is required to describe collective effects in small systems \cite{Grosse-Oetringhaus:2024bwr}.
Taken together,
these experimental difficulties and theoretical deficiencies 
motivate the need for a framework
that simultaneously describes collective effects,
hard-soft correlations,
and jet quenching within a single consistent calculation.
These requirements are especially acute
in small and very peripheral large systems,
where fluctuations at fixed impact parameter
drive centrality selection
and are inextricably coupled to both soft- and hard-particle production,
making independent treatments of each unreliable.
We argue that IP-Glasma~\cite{Schenke:2012wb,Schenke:2012hg} provides
a natural foundation for this unified framework:
IP-Glasma serves as a realistic initial state for subsequent hydrodynamic evolution~\cite{Schenke:2020mbo},
and we will show that IP-Glasma can accurately capture the hard-soft correlations in the kinematic range currently relevant for small-system energy loss.

IP-Glasma~\cite{Schenke:2012wb,Schenke:2012hg} is an initial-state model based on the Color Glass Condensate (CGC)~\cite{Gelis:2010nm}
coupled with realistic nucleon configurations and subnucleonic structure.
The CGC is an effective theory of quantum chromodynamics (QCD) at low Bjorken $x$
and is therefore a natural description of the initial bulk energy density
relevant for low-$p_T$ observables,
where small-$x$ gluons make up most of the degrees of freedom \cite{Gelis:2010nm}.
When used as an initial state
for relativistic viscous hydrodynamic evolution with MUSIC~\cite{Schenke:2010rr,Schenke:2010nt,Schenke:2011bn},
an excellent description of a variety of bulk observables is obtained~\cite{Schenke:2020mbo}.
Notably,
the IP-Glasma + MUSIC framework produced accurate blind predictions
for the ratios of $v_n$ between \oo and \nene collisions~\cite{Mantysaari:2025tcg},
and qualitatively described the $v_n$ in \pp and \ppb collisions~\cite{Schenke:2020mbo},
demonstrating that IP-Glasma provides a realistic initial state
across a wide range of system sizes.

In typical jet quenching calculations~\cite{Wicks:2005gt,Arnold:2002ja,Armesto:2009zi,Schenke:2009gb,JET:2013cls,Casalderrey-Solana:2014bpa,Andres:2016iys,Barata:2020sav,Cao:2020wlm,JETSCAPE:2021ehl,Faraday:2025pto},
hard particle production is computed from pQCD cross sections
via factorization of the partonic cross section and parton distribution functions (PDFs),
while the medium geometry is determined independently---%
decoupling the hard and soft sectors by construction.
Such a construction provides no mechanism for generating correlations
between the hard and soft sectors;
any such correlation must be imposed by hand.
Moreover,
in the momentum range where one expects observable energy loss effects in small systems,
$5~\mathrm{GeV} \lesssim p_T \lesssim 20~\mathrm{GeV}$,
the corresponding Bjorken $x$ range at LHC energies is
$0.001 \lesssim x \lesssim 0.004$---small enough that
$\alpha_s \ln (1/x)$ corrections are potentially large
and the CGC provides a more appropriate effective description \cite{Gelis:2010nm}.
Therefore,
in this work,
we produce all particles---%
both soft and semi-hard---%
from gluons,
as described by IP-Glasma.
Since the soft particles that determine the centrality selection
and the semi-hard particles that probe energy loss
originate from the same event-by-event gluon fields,
hard-soft correlations are naturally included
rather than imposed externally.
Additionally,
we use constraints from a prior Bayesian analysis on HERA data \cite{Mantysaari:2022ffw} 
to constrain the fluctuating proton initial conditions,
meaning that our predictions for hadronic collisions require no further tuning.
While in principle quarks also contribute to charged hadron production at high $p_T$,
these fall outside the IP-Glasma framework.
For $p_T \lesssim 20~\mathrm{GeV}$ at $\sqrt{s_{NN}} = 5.02~\mathrm{TeV}$,
gluons account for $80\text{--}90\%$ of the pion yield~\cite{Sassot:2010bh,Horowitz:2011gd},
and so we anticipate that this omission will not significantly impact our conclusions.

In \cref{sec:model}, we describe the model used to simulate hadronic collisions;
in \cref{sec:multiplicity_distributions,sec:charged_hadron_spectra} we validate our model against data by presenting results for the multiplicity distribution and charged hadron $p_T$ spectra, respectively;
in \cref{sec:nuclear_modification_factor} we show the minimum-bias nuclear modification factor in \po, \ppb, and \oo collisions;
in \cref{sec:centralitydependent_nuclear_modification_factor} we present first results for the centrality-dependent nuclear modification factor from IP-Glasma in \po, \ppb, \oo, and \pbpb collisions;
in \cref{sec:physical_mechanism_that_leads_to_hardsoft_correlations}, we describe the physical mechanism that induces hard-soft correlations in IP-Glasma;
and in \cref{sec:conclusions} we conclude.

\section{Model}
\label{sec:model}

\subsection{IP-Glasma}
\label{sec:ipglasma}

We compute initial conditions for the collisions using the IP-Glasma model \cite{Schenke:2012wb,Schenke:2012hg}, 
which is based on the CGC effective theory of QCD at small Bjorken $x$ \cite{McLerran:1993ka,McLerran:1994vd,Iancu:2003xm}. 
Our implementation follows \cite{Schenke:2020mbo,Mantysaari:2022ffw} 
and utilizes the publicly available \verb|ipglasma| code \cite{Schenke:ipglasma}.

IP-Glasma determines the fluctuating color charge distributions of the colliding nuclei 
and evolves them using classical Yang-Mills (CYM) equations. 
The color charge density is obtained from the IP-Sat model \cite{Kowalski:2003hm}
and encodes both the substructure within each nucleon
and the spatial distribution of nucleons within the nucleus.

In this work we consider protons ($p$), oxygen ions ($\mathrm{O}$), and lead ions ($\mathrm{Pb}$). 
We do not consider any collision system at RHIC, since there the intermediate $p_T$ range probes significantly larger Bjorken $x$,
where our model is not expected to apply.
We assume that protons and neutrons have the same color charge density.

The IP-Sat model \cite{Bartels:2002cj,Kowalski:2003hm} is a dipole model that encodes the impact-parameter dependence of the proton and was fit to inclusive DIS data at HERA \cite{Rezaeian:2012ji}. 
The dipole-proton scattering cross section is parametrized as a function of Bjorken $x$, dipole separation \rt, and impact parameter \bt:
\begin{equation}
    \frac{d\sigma^p_{\rm dip}}{d^2\bt}(x,\rt,\bt) = 2\left[1-\exp\left(-\frac{D(x,\rt,\bt)\,\mathbf{r}_\perp^2}{4}\right)\right]\,,
\end{equation}
where
\begin{equation}
    D(x,\rt,\bt) \equiv \frac{2 \pi^2}{N_c} \alpha_s\left(\tilde{\mu}(\rt)\right) x g(x,\tilde{\mu}^2(\rt)) T_p(\bt),
    \label{eqn:dipole_density_profile}
\end{equation}
with $N_c=3$. In \cref{eqn:dipole_density_profile}, $T_p(\bt)$ denotes the transverse density profile of the proton, and the scale $\tilde{\mu}^2$ is defined as
\begin{equation}
    \tilde{\mu}^2 \equiv \frac{4}{\mathbf{r}_\perp^2}+\tilde{\mu}_0^2\,.
\end{equation}
The running coupling is evaluated at leading order,
\begin{equation}
    \alpha_s(\tilde{\mu}) = \frac{2\pi}{(11-2N_f/3)\ln(\tilde{\mu}/\Lambda_{\rm QCD})}\,,
    \label{eqn:running_coupling_dipole}
\end{equation}
where $N_f=3$ is the number of active quark flavors and $\Lambda_{\rm QCD} = 0.2 ~\mathrm{GeV}$ is the QCD confinement scale.

The gluon density $xg(x,\tilde{\mu}^2)$ at a given value of $x$ is obtained by evolving from the initial scale $\tilde{\mu}_0^2=1.51\,{\rm GeV}^2$
up to $\tilde{\mu}^2$ via leading-order Dokshitzer-Gribov-Lipatov-Altarelli-Parisi (DGLAP) evolution in the approximation that there are no quarks \cite{Rezaeian:2012ji}.
IP-Glasma chooses the initial condition for this evolution to take the form
\begin{equation}
    x g(x,\tilde{\mu}_0^2) = A_g x^{-\lambda_g}(1-x)^{\beta_g}\,,
\end{equation}
following the IP-Sat model \cite{Bartels:2002cj},
where the parameters $A_g=2.308$ and $\lambda_g=0.058$ are fixed by fits to HERA data \cite{Rezaeian:2012ji}. 
The exponent $\beta_g = 5.6$ determines the large-$x$ behavior of the gluon distribution
and is taken from the IP-Sat model \cite{Bartels:2002cj},
where the value was originally motivated by the Martin-Roberts-Stirling-Thorne (MRST)
parameterization of the gluon density \cite{Martin:2001es}.

The saturation radius $\mathbf{r}_s$ is defined in terms of the density profile $D$ from \cref{eqn:dipole_density_profile} as
\begin{equation}
  \exp\left( -\frac{D(x, \mathbf{r}_s, \bt) \mathbf{r}_s^2}{2}\right) \equiv e^{-1}.
  \label{eqn:saturation_radius}
\end{equation}
The saturation scale $Q_s$ is then defined as
\begin{equation}
  Q_s^2(x, \bt) \equiv D(x, \mathbf{r}_s, \bt).
  \label{eqn:saturation_scale}
\end{equation}
\Cref{eqn:saturation_radius,eqn:saturation_scale} together imply that $Q_s^2 = 2 / \mathbf{r}_s^2$.

IP-Glasma includes an event-by-event fluctuating density by writing the proton transverse density profile as
\begin{equation}
T_p\left(\mathbf{b}_{\perp}\right) \equiv \frac{1}{N_q} \sum_{i=1}^{N_q} p_i T_q\left(\mathbf{b}_{\perp}-\mathbf{b}_{\perp, i}\right),
\label{eqn:proton_thickness_function}
\end{equation}
where
\begin{equation}
T_q\left(\mathbf{b}_{\perp}\right) \equiv \frac{1}{2 \pi B_q} e^{-\mathbf{b}_{\perp}^2 /\left(2 B_q\right)},
\label{eqn:quark_thickness_function}
\end{equation}
$N_q$ is the number of hotspots with RMS radius $\sqrt{2 B_q}$,
and $B_q$ is the ``hotspot size'' parameter sampled from the Bayesian posterior from fitting to HERA data \cite{Mantysaari:2022ffw};
we will discuss the details of the Bayesian posterior sampling in \cref{sec:bayesian_posterior_sampling}.
We take $N_q = 3$,
corresponding to the three valence quarks in the proton,
following previous work~\cite{Schenke:2012wb,Schenke:2012hg,Schenke:2013dpa,Mantysaari:2016ykx,Schenke:2020mbo}.
While $N_q$ can in principle be treated as a free parameter---%
with deviations from three reflecting emergent large-$x$ subnucleonic structure
beyond the valence quarks---%
$N_q = 3$ is a natural choice
and consistent with both HERA data \cite{Mantysaari:2016ykx,Mantysaari:2022ffw}
and heavy-ion data following hydrodynamic evolution \cite{Schenke:2020mbo}.
However,
we caution the reader against identifying the $N_q = 3$ hotspots
with the valence quarks themselves:
the root-mean-square gluonic radius extracted in \cite{Mantysaari:2022ffw},
$r^{g}_{\text{RMS}} = \num{0.610(65:68)}~\mathrm{fm}$,
is significantly smaller than the measured proton charge radius,
$r^{\pm}_{\text{RMS}} = 0.8409 \pm 0.0004~\mathrm{fm}$ \cite{ParticleDataGroup:2022pth}.

Each hotspot center $\bt^{i}$ is sampled from a distribution
\begin{equation}
  P(\bt^{i}) = \frac{1}{2 \pi B_{qc}} e^{-\left(\bt^{i}\right)^2 / \left(2 B_{qc}\right)},
  \label{eqn:hotspot_location_distribution}
\end{equation}
where $B_{qc}$ is a parameter sampled from the HERA Bayesian posterior \cite{Mantysaari:2022ffw}. 
Motivated by possible short-range correlations between hotspots,
a minimum distance $d_{q,\text{min}}$ is introduced,
which will also be sampled from the Bayesian posterior.

IP-Glasma additionally includes saturation scale fluctuations
by allowing the density of each hotspot to fluctuate independently.
The dimensionless weights $p_i$ appearing in \cref{eqn:proton_thickness_function}
are sampled from a log-normal distribution
\begin{equation}
  P\left(\ln p_i\right)
  = \frac{1}{\sqrt{2 \pi} \sigma} \exp\left[-\frac{\ln^2 p_i}{2 \sigma^2}\right],
  \label{eqn:saturation_scale_fluctuations}
\end{equation}
with one $p_i$ per hotspot.
The sampled $p_i$ are then divided by the expectation value of this distribution,
$E[p] = e^{\sigma^2 / 2}$,
so that the fluctuations leave the average density unmodified~\cite{Mantysaari:2022ffw}.
The width $\sigma$ is sampled from the HERA Bayesian posterior.

For extended nuclei,
the positions of the individual nucleons must additionally be sampled.
Oxygen nuclei are described by nucleon configurations obtained from a variational
Monte Carlo (VMC) calculation using the Argonne $v_{18}$ (AV18) two-nucleon potential
with Urbana X (UX) three-nucleon interactions \cite{Carlson:1997qn}.
We use the tabulated configurations distributed with IP-Glasma, for which
nucleon-nucleon correlations, including the short-range repulsion, are inherited
from the many-body wave function and no additional minimum-distance cut is applied.
For lead, nucleon positions are sampled from a Woods-Saxon distribution
\begin{equation}
\rho(r) \propto \frac{1}{1+\exp \left[\left(r-R\right) / a\right]}
  \label{eqn:woods_saxon}
\end{equation}
with $R = 6.62 ~\mathrm{fm}$ and $a = 0.546 ~\mathrm{fm}$~\cite{DeVries:1987atn}, imposing a minimum
inter-nucleon distance $d_{\text{min}} = 0.9 ~\mathrm{fm}$~\cite{Schenke:2020mbo}.
Beyond this hard core, no short-range correlations are included for lead.

Collisions $A+B$ are generated by sampling two random nucleon configurations, independently rotating each nucleus by a uniformly distributed random orientation, and sampling the impact parameter $\mathbf{b}$ uniformly in the transverse plane.
Collisions that do not have any overlapping color charge density are not included in the set of ``minimum-bias'' collisions. 
However, we do keep track of the total number of events with no overlapping color charge density for later computation of the inelastic cross section; see \cref{sec:observable_implementation}.
All nucleons are first sampled,
then the density profile of each nucleon is determined according to \cref{eqn:proton_thickness_function},
with hotspot locations sampled from \cref{eqn:hotspot_location_distribution}. 
The total density profile of the nucleus is then computed by summing over the contributions from each nucleon, 
$T_A(\bt) \equiv \sum_j T_{p,j}(\bt)$.

Since the local saturation scale $Q_s(x, \mathbf{b}_{\perp})$ depends on $x$,
$Q_s(x, \bt)$ must be determined iteratively via the relationship $x = Q_s(x, \mathbf{b}_{\perp}) / \sqrt{s_{NN}}$,
which is valid at midrapidity. 
IP-Glasma is a rapidity-independent framework,
and we evaluate all observables in a single slice at $y = 0$.
We therefore make no attempt to describe the rapidity dependence of particle production,
which is substantial in asymmetric systems such as \ppb,
and compare only to midrapidity data throughout.
Color charges are then sampled under the assumption of local Gaussian correlations,
as in the McLerran-Venugopalan (MV) model \cite{McLerran:1993ka, McLerran:1994vd},
with mean zero and variance
\begin{equation}
  \left\langle\rho_l^a\left(\mathbf{b}_{\perp}\right) \rho_l^b\left(\mathbf{x}_{\perp}\right)\right\rangle=g^2 \mu_l^2\left(x, \mathbf{b}_{\perp}\right) \delta^{a b} \delta^{(2)}\left(\mathbf{b}_{\perp}-\mathbf{x}_{\perp}\right),
  \label{eqn:mv_model_variance}
\end{equation}
where $l = A, B$ labels the two colliding nuclei, which are sampled independently,
and $a$ and $b$ are adjoint color indices.
Here $g^2 \mu_l\left(x, \mathbf{b}_{\perp}\right)=C Q_{s,l}\left(x, \mathbf{b}_{\perp}\right)$ for a proportionality constant $C$,
where $Q_{s,l}$ is the local saturation scale of nucleus $l$ from \cref{eqn:saturation_scale}. The value of $C$ will be sampled from the HERA Bayesian posterior.
The coupling constant $g$ scales out of all classical calculations 
and only affects the normalization of observables like the gluon number.
We will take $g = 1$ for all simulations, consistent with previous work \cite{Schenke:2013dpa},
and restore the coupling at the end of the calculation
through the overall factor $g^2 / (4\pi \alpha_s(\tilde{\mu}))$;
the details are discussed at the end of this section.

The color charges from both projectile and target are used to construct pure gauge configurations for each nucleus before the collision. 
The initial conditions for the gauge fields immediately after the collision are obtained by solving a set of matching conditions at $\tau = 0^+$ \cite{Krasnitz:1998ns,Schenke:2012wb}. 
Under the assumption of boost invariance, the fields are then evolved in proper time $\tau$ using the CYM equations of motion on a 2D transverse lattice via a leapfrog algorithm. 
A detailed description of the lattice formulation and equations of motion can be found in \cite{Schenke:2012wb,Schenke:2012hg,Schenke:2013dpa,Schenke:2020mbo}. 

For \pbpb collisions we used a lattice size of $L=30\,{\rm fm}$,
for \oo collisions we used $L=15 ~\mathrm{fm}$,
for \pa collisions we used $L = 12 ~\mathrm{fm}$,
and for \pp collisions we used $L = 6 ~\mathrm{fm}$.
In this work
we are interested in high-$p_T$ physics,
and therefore must consider smaller lattice spacings
than are typical for works using IP-Glasma as an initial condition for hydrodynamics.
We take a lattice spacing $a = 0.015 ~\mathrm{fm} \sim 0.076 ~\mathrm{GeV}^{-1}$ for all systems, which enables reasonable predictions with minimal lattice effects out to around $p_T \sim (\pi / 2) / a \sim 21 ~\mathrm{GeV}$.
Subsequent hadronization pushes the onset of the lattice artifacts to lower $p_T$ than this estimate; however, we show in 
\cref{sec:dependence_of_results_on_numerical_choices} that our main results are largely insensitive to both the lattice spacing $a$ and the box size $L$.

The number of gluons can be written on the lattice 
in units of $a = 1$ (lattice units) as
\begin{equation}
\begin{aligned}
  \frac{d^3 N_g}{dk_x\, dk_y\, dy}= \frac{2}{N^2} \frac{1}{\tilde{k}_T}\bigg[ & \frac{g^2}{\tau} \operatorname{Tr}\left(E_i\left(\mathbf{k}_{\perp}\right) E_i\left(-\mathbf{k}_{\perp}\right)\right) \\
& +\tau \operatorname{Tr}\left(\pi\left(\mathbf{k}_{\perp}\right) \pi\left(-\mathbf{k}_{\perp}\right)\right)\bigg],
\end{aligned}
\label{eqn:gluon_multiplicity_lattice_units}
\end{equation}
with $N$ the number of lattice sites in one transverse dimension,
$k_x$ and $k_y$ the $x$- and $y$-components of the gluon momentum in lattice units,
$\mathbf{k}_{\perp} \equiv (k_x, k_y)$,
$E_i$ the transverse chromoelectric field components,
and $\pi \equiv E^{\eta}$ the longitudinal chromoelectric field component,
canonically conjugate to $A^{\eta}$.
The dispersion relation on the lattice in lattice units yields
\begin{equation}
\tilde{k}_T^2=4\left[\sin ^2 \frac{k_x}{2}+\sin ^2 \frac{k_y}{2}\right],
\label{eqn:dispersion_relation}
\end{equation}
which is the effective lattice momentum squared. To obtain the gluon multiplicity in physical units, one simply includes the appropriate factors of the lattice spacing $a$,
\begin{equation}
  \frac{d^3 N_g}{dp_x\, dp_y\, dy} = a^2 \frac{d^3 N_g}{dk_x\, dk_y\, dy},
  \label{eqn:gluon_multiplicity_physical_units}
\end{equation}
where $p_i \equiv k_i / a$ is the physical momentum. 

As noted above, 
solutions to the Yang-Mills equations carry no dependence on
the coupling constant $g$,
which enters only as an overall multiplicative factor in the final multiplicity.
Both terms in the square brackets in \cref{eqn:gluon_multiplicity_lattice_units} are proportional to $1/g^2$ \cite{Schenke:2013dpa}.
Therefore, the effects of running coupling on the gluon multiplicity can be included by multiplying \cref{eqn:gluon_multiplicity_physical_units} by $g^2/(4\pi \alpha_s(\tilde{\mu}))$.
In this work,
we take $\tilde{\mu}=p_T/2$, consistent with previous work~\cite{Schenke:2013dpa}, 
where $p_T = | \mathbf{p}_T|$ is the transverse gluon momentum in physical units. 
In \cref{sec:dependence_of_results_on_physical_choices} we compare against two common alternatives:
switching the running off entirely,
and evaluating the coupling at $\tilde{\mu} = Q_s^{\text{max}}$,
the larger of the two saturation scales in the collision.

We use the regulated one-loop prescription for the running coupling 
\begin{equation}\label{eq:running}
  \alpha_s(\tilde{\mu}) = \frac{4\pi}{\beta \ln\left(\left[(\mu_0/\Lambda_{\rm QCD})^{2/c} + (\tilde{\mu}/\Lambda_{\rm QCD})^{2/c}\right]^c\right)}\,.
\end{equation}
The quantity $\mu_0$ regulates the pole 
and is taken as $\mu_0=0.5\,{\rm GeV}$ following previous work \cite{Schenke:2013dpa}.
The parameter $c$ controls the sharpness of the cutoff and is taken as $c = 0.2$, consistent with previous work \cite{Schenke:2013dpa}.
It has been shown that varying $c$ does not significantly change the results \cite{Schenke:2013dpa}.
Varying $\mu_0$ leads to changes in the absolute value of the multiplicity,
which can be absorbed into changes in $g$,
but does not change the shape of the multiplicity distribution~\cite{Schenke:2013dpa}.
We take $N_c = 3$, $\beta = 11 - 2 N_f /3$, $N_f = 3$, and $\Lambda_{\text{QCD}} = 0.2 ~\mathrm{GeV}$ throughout.

We note that two distinct running coupling prescriptions appear in our calculation,
and that they are consistent with one another.
\Cref{eqn:running_coupling_dipole} is evaluated at the dipole scale
and is internal to the IP-Sat fit to HERA data \cite{Rezaeian:2012ji},
which we retain unmodified;
\cref{eq:running} is evaluated at the scale of the produced gluon
and enters only in the conversion of the classical field to a gluon number.
The two are the same one-loop coupling,
each evaluated at the scale of the corresponding process,
and differ only in how the pole is regulated:
\cref{eq:running} reduces to \cref{eqn:running_coupling_dipole}
for $\tilde{\mu} \gg \mu_0$.

The gluon spectrum in \cref{eqn:gluon_multiplicity_physical_units} is then hadronized using fragmentation functions
\begin{equation}
  \frac{d^3 N^h}{d^2 p_T^{h}\, dy}\left(p_T^{h}\right)=
  \int \frac{dz}{z^2} \frac{d^3 N^g}{d^2 p_T^{g}\, dy}\left(\frac{p_T^h}{z}\right) D_g^h(z, Q),
  \label{eqn:hadron_spectrum}
\end{equation}
where $p_T^g \equiv p_T^h / z$ is the physical transverse gluon momentum,
$D_g^h(z, Q)$ is the fragmentation function for producing a hadron $h$ from a gluon $g$,
$Q = p_T^g$ is the energy scale,
and $p_T^h$ is the hadron momentum.
As discussed in the Introduction, 
quarks are not modeled in the IP-Glasma framework; however,
since gluons account for $80\text{--}90\%$ of the pion yield for $p_T \lesssim 20~\mathrm{GeV}$ at $\sqrt{s_{NN}} = 5.02~\mathrm{TeV}$~\cite{Sassot:2010bh,Horowitz:2011gd},
we anticipate that neglecting quarks will not significantly impact our conclusions.
We emphasize that, because the overall normalization of the spectrum is fixed by hand
(see \cref{sec:charged_hadron_spectra}),
the omitted quark contribution is effectively absorbed into that normalization.
In this work we use charged hadron fragmentation functions from the de Florian, Sassot, and Stratmann (DSS) set \cite{deFlorian:2007aj}; we assess the impact of the choice of fragmentation function in \cref{sec:dependence_of_results_on_physical_choices}.

\subsection{Bayesian posterior sampling and uncertainties}
\label{sec:bayesian_posterior_sampling}

All results in this work use IP-Glasma parameters
sampled from the posterior of a Bayesian analysis~\cite{Mantysaari:2022ffw}.
In~\cite{Mantysaari:2022ffw}, the authors performed a Bayesian analysis
of ZEUS and H1 measurements of coherent and incoherent diffractive $J/\psi$ production
in electron-proton collisions at HERA
to constrain the IP-Glasma parameters describing the fluctuating proton.
The analysis was performed both with $N_q = 3$ hotspots
and with $N_q$ treated as a free parameter.
The free-$N_q$ analysis found that $N_q \geq 2$ is required
to describe the HERA data,
but could not further constrain $N_q$.
In this work we use the $N_q = 3$ posterior for simplicity.

The free parameters of the analysis,
together with their maximum \emph{a posteriori} (MAP) values,
i.e.\ the most likely values, with uncertainty estimates given as 90\% confidence intervals,
are listed in \cref{tab:map_parameters}.

\begin{table}[tb]
  \caption{Free parameters of the Bayesian analysis of \cite{Mantysaari:2022ffw} for $N_q = 3$,
  with their MAP values and 90\% confidence intervals.}
  \label{tab:map_parameters}
  \begin{ruledtabular}
  \renewcommand{\arraystretch}{1.35}
  \begin{tabular}{llc}
    Parameter & Description & MAP value \\
    \colrule
    $m$ & Infrared regulator & \num{0.246(162:103)} $\mathrm{GeV}$ \\
    $B_{qc}$ & Proton size & \num{4.45(80:80)} $\mathrm{GeV}^{-2}$ \\
    $B_q$ & Hotspot size & \num{0.346(282:202)} $\mathrm{GeV}^{-2}$ \\
    $\sigma$ & Size of $Q_s$ fluctuations & \num{0.563(143:141)} \\
    $C^{-1}$ & Ratio $Q_s / (g^2 \mu)$ & \num{0.747(070:093)} \\
    $d_{q,\text{min}}$ & Minimum hotspot distance & \num{0.254(222:229)} $\mathrm{fm}$ \\
  \end{tabular}
  \end{ruledtabular}
\end{table}

The MAP values do not capture the correlations between the preferred values of the parameters, 
and so we sample from the posterior to obtain an estimate of the systematic uncertainty arising from uncertainties in the extracted parameters.

We sample 20 parameter sets from the posterior
and run IP-Glasma for each of them,
generating ${\sim}5000$ events per collision system per parameter set. 
All systems entering a given observable
are generated with the same parameter set,
so combinations of systems such as $R_{AB}$ are formed sample by sample.
Correlations between systems are therefore retained exactly,
and systematic shifts common to both the numerator and denominator of $R_{AB}$ cancel.
The systematic $68\%$ confidence
interval is the $16$th--$84$th percentile range of the resulting distribution,
which is shown as a band in most figures.
For \ppb we checked that increasing the number of parameter sets
from 20 to 80 did not change our results.

We additionally show statistical uncertainties arising from the finite number of simulated events,
estimated by bootstrap resampling the events of every system within each parameter set. 
Because the \ab and \pp event samples are statistically independent, their statistical uncertainties do not cancel in observables such as $R_{AB}$ that combine the two systems.
We report the average of these uncertainties over all parameter sets in the figures.
We note that since each posterior variation is evaluated on a finite sample of events, 
the observed spread in the posteriors contains a statistical component in addition to the genuine systematic variation. 
The quoted systematic uncertainty should therefore be read as an upper bound on the true systematic,
especially when the statistical uncertainty is comparable in size to the systematic uncertainty.

\subsection{Observables}
\label{sec:observable_implementation}

In this work we focus on three classes of observable: the charged hadron spectrum $d^3 \Nch / d^2 p_T\, dy$, the self-normalized multiplicity distribution $P(\Nch / \avg{\Nch})$, and the nuclear modification factor $R_{AB}$ for the collision \ab. Before turning to the implementation of each observable, we first discuss the determination of centrality.

Centrality is defined in several related ways.
Experimentally, centrality is usually taken as percentiles of the multiplicity or energy deposited in a particular detector:
CMS and ALICE use the charged-particle multiplicity at forward rapidity \cite{ALICE:2013hur},
while ATLAS uses the forward transverse energy \cite{ATLAS:2022kqu}.
Forward rapidity is preferred experimentally because a forward estimator reduces the contamination of the centrality determination by jets at midrapidity,
which are often the subject of measurements.

In this work, the framework is evaluated entirely at midrapidity $y=0$,
and so we cannot match the standard experimental definition of centrality.
We define centrality, therefore, 
as percentiles of the total multiplicity at $y=0$.
This choice is standard in the field \cite{Bernhard:2019bmu,Schenke:2020mbo,Nijs:2020roc},
where $(2+1)$D hydrodynamic simulations, boost invariant by construction, are most common.
We anticipate that this approximation will be reasonable for large symmetric collision systems,
which are approximately boost invariant \cite{Bjorken:1982qr}.

Centrality is significantly more complicated in small collision systems,
where there are large hard-soft correlations and selection biases.
To reduce such biases in experimental measurements,
it is common to use the energy deposited by spectator nucleons at zero degrees (down the beam pipe) as a centrality measure \cite{ALICE:2014xsp}.
Recreating such a centrality definition requires detailed modeling both of the breakup of the incoming nucleus after the collision and of the coalescence of the spectators into nuclear fragments,
and is therefore beyond the scope of this work. 
It would be interesting in the future to study what, if any, 
hard-soft correlations remain in events 
where the centrality is selected using zero-degree spectators \cite{Broomhall:2026}.
In this work we will only compare our centrality-selected small-system results
to data from ALICE \cite{ALICE:2014xsp},
where the nuclear modification factor was reported for various definitions of centrality.
In particular, the results selected according to the ``CL1'' prescription define centrality using the multiplicity within $|\eta| < 1.4$,
making them an appropriate point of comparison for our midrapidity results.

We will now describe how we compute each observable presented in this work.

The charged hadron spectrum is calculated according to \cref{eqn:hadron_spectrum}.
Our spectrum is differential in rapidity,
while experiments report the spectrum differential in pseudorapidity, $d^3 \Nch / d^2 p_T\, d\eta$.
At the physical pion mass the Jacobian relating the two is $dy / d\eta = p_T / E$ at $y = 0$,
which differs from unity by $\lesssim 1\%$ for $p_T \gtrsim 1 ~\mathrm{GeV}$.
Over the $5\text{--}20 ~\mathrm{GeV}$ range in which we compare spectra,
the two are therefore interchangeable, and we compare them directly.
The Jacobian is not negligible at the lower edge of the $p_T$ window used for the
multiplicity distributions, reaching $\simeq 0.73$ at $p_T = 0.15 ~\mathrm{GeV}$;
however, as a fixed function of $p_T$ independent of event activity,
the Jacobian acts as a common rescaling of the soft yield
and should largely cancel in the self-normalized $P(\Nch / \avg{\Nch})$.
The same argument applies to the centrality determination,
which depends only on percentiles of the multiplicity distribution and is therefore invariant under any monotonic common rescaling.
We do not apply the Jacobian, since doing so would reinstate a finite hadron mass that is set to zero throughout \cref{eqn:hadron_spectrum}, including in the extraction of the fragmentation functions we employ \cite{deFlorian:2007aj}, and would therefore be a partial correction of the same order as terms neglected elsewhere.

The self-normalized multiplicity distribution is calculated as follows. 
For each event,
the total number of charged particles within the $p_T$ range $[p_{T, \text{min}}, p_{T, \text{max}}]$ specified by the particular experiment is calculated as
\begin{equation}
  \Nch^{i} \equiv \int_{p_{T, \text{min}}}^{p_{T, \text{max}}} d p_T\; 2 \pi p_T \frac{d^3 \Nch}{d^2 p_T\, dy}
  \label{eqn:nch}
\end{equation}
from \cref{eqn:hadron_spectrum} for each event $i$.
Since \cref{eqn:hadron_spectrum} is differential in rapidity,
$\Nch$ as defined in \cref{eqn:nch} is a midrapidity density, $d\Nch/dy$ at $y = 0$;
the rapidity window is common to every event and cancels in the self-normalized distribution below.
We then compute the average over all events
\begin{equation}
  \avg{\Nch} \equiv \frac{1}{N_{\text{ev}}} \sum_{i=1}^{N_{\text{ev}}} \Nch^{i}
  \label{eqn:average_nch}
\end{equation}
and histogram $\Nch / \avg{\Nch}$ over all events, which we denote $P(\Nch / \avg{\Nch})$.

The nuclear modification factor for the collision system \ab in some event selection (typically a centrality class) is defined as
\begin{equation}
  R_{AB} (p_T) \equiv \frac{1}{\avg{\Ncoll}} \frac{d^3 \Nch^{AB} / d^2 p_T\, dy}{d^3 \Nch^{pp} / d^2 p_T\, dy},
  \label{eqn:nuclear_modification_factor}
\end{equation}
where $\avg{\Ncoll}$ is the average number of binary nucleon-nucleon collisions for the given event selection,
$d^3 \Nch^{AB} / d^2 p_T\, dy$ is the charged hadron spectrum for the collision system \ab,
and $d^3 \Nch^{pp} / d^2 p_T\, dy$ is the charged hadron spectrum for minimum-bias \pp collisions.
Although $R_{AB}$ is an experimental observable,
the measurement requires theoretical input in the form of $\Ncoll$ for each centrality class.

For minimum-bias collisions, a model-independent form of the nuclear modification factor can be constructed
\begin{equation}
  R_{AB}^{\sigma} \equiv \frac{1}{AB} \frac{d^3 \sigma_{\text{ch}}^{AB} / d^2 p_T\, dy}{d^3 \sigma_{\text{ch}}^{pp} / d^2 p_T\, dy},
  \label{eqn:nuclear_modification_factor_sigma}
\end{equation}
where
$d^3 \sigma_{\text{ch}}^{AB} / d^2 p_T\, dy$ is the charged hadron production cross section in \ab collisions,
and $d^3 \sigma_{\text{ch}}^{pp} / d^2 p_T\, dy$ is the charged hadron production cross section in \pp collisions.
While the main focus of this work will be on centrality-selected collision systems, we will utilize \cref{eqn:nuclear_modification_factor_sigma}
to compare to recent experimental \oo and \po data \cite{ALICE:2026zck,CMS:2025bta},
and to understand the overall normalization and asymptotic behavior of our results.
The inelastic cross section may be computed in IP-Glasma using 
\begin{align}
  \sigma_{\text{inel}}^{AB} &= \int_{0}^{b_{\text{max}}} db \; 2 \pi b \; P_{\text{inel}}^{AB}(b) \\
  &= \pi b^2_{\text{max}} \int_{0}^{b_{\text{max}}} db \; \frac{2 b}{b^{2}_{\text{max}}} \; P_{\text{inel}}^{AB}(b) \\
  &= \pi b_{\text{max}}^2 \left\langle P_{\text{inel}}^{AB} \right\rangle \\
  &\simeq \pi b_{\text{max}}^2 \frac{N_{\text{inel}}}{N_{\text{ev}}} \, ,
  \label{eqn:inelastic_cross_section}
\end{align}
where $b_{\text{max}}$ is the maximum impact parameter sampled in the simulation, $P_{\text{inel}}^{AB}(b)$ is the probability of an inelastic interaction at impact parameter $b$, $N_{\text{inel}}$ is the number of inelastic events, and $N_{\text{ev}}$ is the total number of simulated events.
For each collision system we choose $b_{\text{max}}$ large enough that $P_{\text{inel}}^{AB}(b_{\text{max}}) \simeq 0$. 
Since events are sampled uniformly in $b^2$ on $[0, b_{\text{max}}^2]$,
$N_{\text{inel}} / N_{\text{ev}}$ is an unbiased estimator of $\left\langle P_{\text{inel}}^{AB} \right\rangle$.

We define an inelastic event as one in which at least one lattice cell receives a nonzero color
charge density from both nuclei.
This condition requires the local thickness of each nucleus at that cell to exceed a threshold,
which in our implementation is the lower edge of the tabulated $Q_s^2(T_p)$ grid,
$T_p \geq 10^{-4} ~\mathrm{GeV}^2$, corresponding to $Q_s \gtrsim 40 ~\mathrm{MeV}$;
we discuss the consequences of this choice in \cref{sec:inelastic_cross_section_parameters}.
One may then write
\begin{equation}
  \frac{d^3 \sigma_{\text{ch}}^{AB}}{d^2 p_T\, dy} = \sigma_{\text{inel}}^{AB} \; \frac{d^3 \Nch^{AB}}{d^2 p_T\, dy}.
  \label{eqn:differential_cross_section}
\end{equation}
One anticipates that $R_{AB} \simeq R^{\sigma}_{AB}$.
The two agree, however, only to the extent that the $\sigma_{\text{inel}}^{NN}$ used in the Glauber determination of \Ncoll agrees with that produced by IP-Glasma;
we quantify the difference in \cref{sec:nuclear_modification_factor},
and derive the exact relation between the two forms in \cref{sec:inelastic_cross_section_parameters}.

\Cref{eqn:nuclear_modification_factor} is the only form of $R_{AB}$ that permits centrality selection, and requires a model for $\Ncoll$.
Experimentalists~\cite{CMS:2011iwn,ATLAS:2011ag,ALICE:2013hur} typically use
a Monte Carlo implementation of the Glauber model~\cite{Miller:2007ri},
or in some cases a variant that includes Glauber--Gribov corrections \cite{Guzey:2005tk,Alvioli:2013vk,ATLAS:2016xpn},
to calculate $\Ncoll$ in each centrality class.
The standard procedure for determining centrality
is to run a Monte Carlo Glauber model for random collisions
and collect pairs $(\Ncoll, N_{\text{part}})$,
where $N_{\text{part}}$ is the number of participating nucleons.
A binary collision is counted for every inelastic nucleon-nucleon collision,
with the inelastic cross section $\sigma_{\text{inel}}^{NN}$ as an input parameter.
A participant is counted once for every nucleon
that undergoes at least one inelastic binary collision.
One then fits a negative binomial distribution (NBD)
convolved with the distribution of $N_{\text{part}}$,
or sometimes with a linear combination of $\Ncoll$ and $N_{\text{part}}$ \cite{Kharzeev:2000ph,ALICE:2013hur},
to the measured charged-particle multiplicity distribution $P(\Nch)$
in a particular detector.
Applying the same centrality cuts to the NBD $\otimes$ Glauber model output
then yields $\Ncoll$ for each centrality class in a data-driven way.

When $R_{AB}$ is computed theoretically,
the result is often largely insensitive to the number of binary collisions.
In energy loss calculations, for example, one constructs
\begin{equation}
    \frac{d^3 N_{AB}}{d^2 p_T\, dy}
    = \Ncoll \left\langle \frac{d^3 N_{ab}}{d^2 p_T\, dy} \right\rangle ,
\end{equation}
where $a$ and $b$ are nucleons sampled from $A$ and $B$,
and the average runs over production points
drawn from the normalized binary collision density
\cite{Cao:2020wlm,Faraday:2025pto}.
Here $d^3 N_{ab} / d^2 p_T\, dy$ is not the free nucleon-nucleon yield:
rather,
$d^3 N_{ab} / d^2 p_T\, dy$ includes,
in principle,
cold nuclear matter effects such as nPDF effects and hot nuclear matter effects such as energy loss.
The ratio $R_{AB}$ is therefore by construction independent of \Ncoll.
In this work, we make no such \emph{a priori} assumption
of proportionality to $\Ncoll$.
Instead,
when computing $R_{AB}$,
we use the $d^3 \Nch^{AB} / d^2 p_T\, dy$ and $d^3 \Nch^{pp} / d^2 p_T\, dy$ directly from IP-Glasma
for the \ab and \pp collisions, respectively,
together with the experimentally reported $\Ncoll$ for the corresponding centrality class.
Because we adopt the experimentally reported $\Ncoll$,
the normalization uncertainty on $\Ncoll$ is shared between our results and the data,
and we therefore do not propagate this normalization uncertainty to our theoretical results.
When the experimentally reported $\Ncoll$ is not available,
for instance in \oo collisions, where the centrality-cut $R_{AA}$ has not yet been measured,
we shall use a value computed as follows.
$\Ncoll$ is not used in IP-Glasma;
we compute \Ncoll by running a Monte Carlo Glauber model
on the same initial nucleon configurations.
We take the inelastic nucleon-nucleon cross section as $\sigma_{\text{inel}}^{NN} = 67~\mathrm{mb}$, consistent with the parameterization of Ref.~\cite{dEnterria:2020dwq}.

\begin{figure}[!th]
  \centering
  \includegraphics[width=\linewidth]{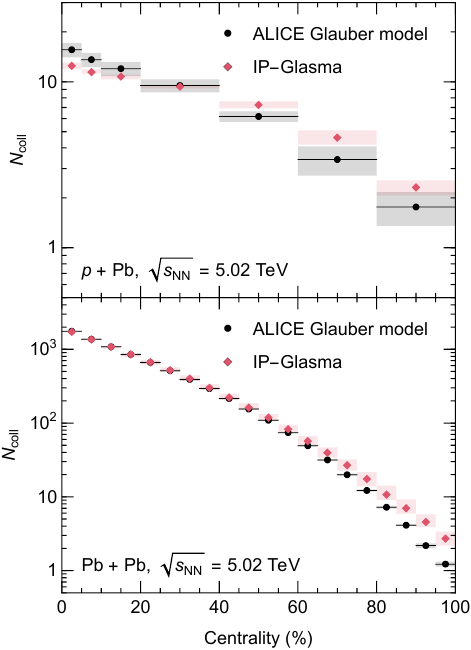}
  \caption{Number of binary collisions $\Ncoll$ as a function of centrality for $\sqrt{s_{NN}} = 5.02 ~\mathrm{TeV}$ \ppb (top) and $\sqrt{s_{NN}} = 5.02 ~\mathrm{TeV}$ \pbpb (bottom) collisions. IP-Glasma results are shown in red and results from the ALICE collaboration~\cite{ALICE:2014xsp,ALICE:2018ekf} are shown in black. Red boxes are systematic theoretical uncertainties and gray boxes are systematic experimental uncertainties. Statistical uncertainties are smaller than the marker size and are not visible. Horizontal black lines represent the size of the centrality bin.}
  \label{fig:ncoll_combined_vs_centrality}
\end{figure}

We will briefly investigate the importance of the decision
to use the experimentally reported $\Ncoll$
by comparing the values from IP-Glasma to those reported by ALICE.
\Cref{fig:ncoll_combined_vs_centrality} plots the number of binary collisions $\Ncoll$ as a function of centrality 
for $\sqrt{s_{NN}} = 5.02 ~\mathrm{TeV}$ \ppb collisions (top) and
$\sqrt{s_{NN}} = 5.02 ~\mathrm{TeV}$ \pbpb collisions (bottom). 
Experimentally reported values by the ALICE collaboration are shown in black~\cite{ALICE:2014xsp,ALICE:2018ekf}.
The centrality used for the experimental $\Ncoll$ in \ppb collisions is defined by the CL1 estimator, i.e., the multiplicity within $|\eta| < 1.4$, which is the most comparable to our IP-Glasma results. Since such a centrality determination was not available for \pbpb, the centrality used for the experimental $\Ncoll$ in \pbpb collisions is determined from the V0M estimator,
i.e., the summed multiplicity in $2.8 < \eta < 5.1$ and $-3.7 < \eta < -1.7$.

We observe in the top panel of \cref{fig:ncoll_combined_vs_centrality}
that the Glauber-model $\Ncoll$
associated with the nucleon configurations used by IP-Glasma
is systematically smaller in central \ppb collisions
than the $\Ncoll$ reported by ALICE,
and systematically larger in peripheral \ppb collisions.
The bottom panel of \cref{fig:ncoll_combined_vs_centrality}
shows the same qualitative trend in \pbpb collisions.
Because $\Ncoll$ is so large in central \pbpb collisions, however,
the relative difference there is small,
and the effect is visible only in peripheral collisions.
The systematic difference in the centrality dependence of $\Ncoll$
between the experimental Glauber model and IP-Glasma
motivates our use of the experimentally reported values of \Ncoll throughout this work,
which avoids introducing a non-physical centrality bias into the theoretical $R_{AB}$;
using instead the $\Ncoll$ from IP-Glasma
would increase the centrality dependence of our reported $R_{AB}$.

\begin{figure*}[!t]
  \centering
  \includegraphics[width=\linewidth]{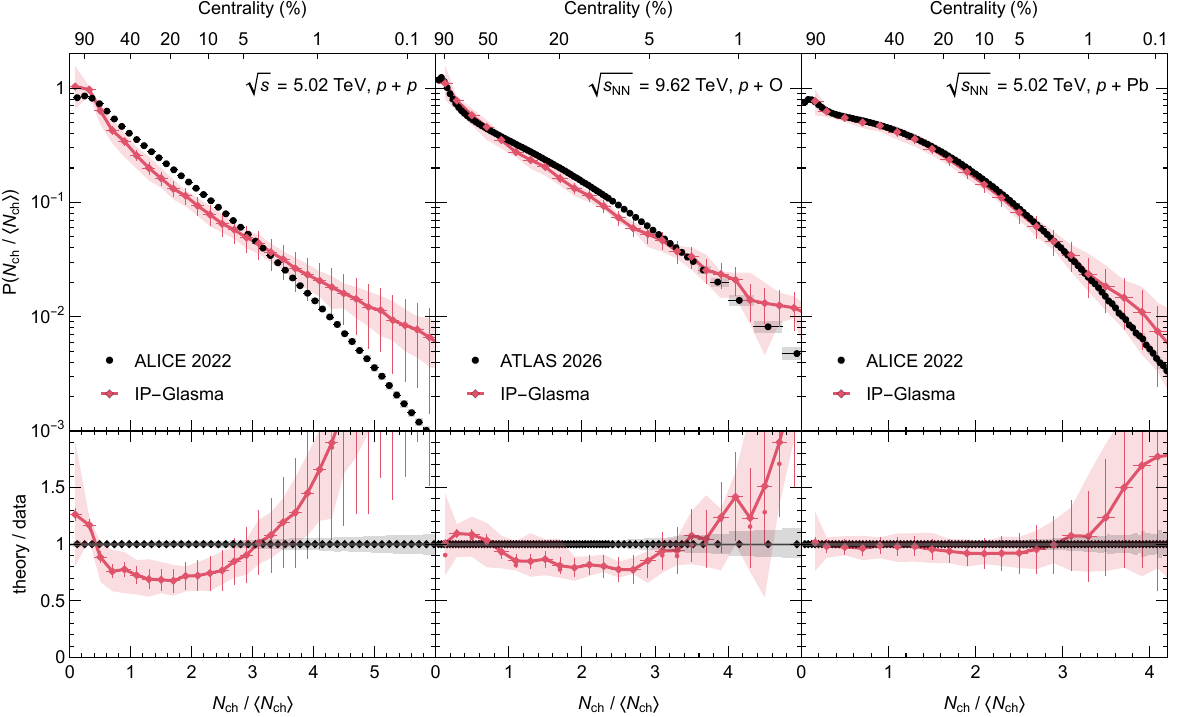}
  \caption{The self-normalized multiplicity distribution $P(\Nch / \avg{\Nch})$ for charged hadrons in $\sqrt{s} = 5.02~\mathrm{TeV}$ \pp (left), $\sqrt{s_{NN}} = 9.62~\mathrm{TeV}$ \po (center), and $\sqrt{s_{NN}} = 5.02~\mathrm{TeV}$ \ppb (right) collisions. Top row: $P(\Nch / \avg{\Nch})$ as a function of $\Nch / \avg{\Nch}$, with the corresponding centrality indicated on the upper axis. Bottom row: ratio of theory to experimental data. IP-Glasma results are shown in red and experimental data in black: ALICE~\cite{ALICE:2022xip} for \pp and \ppb, and ATLAS~\cite{ATLAS:2026xcm} for \po, converted from the reported $d N_{\text{ev}} / d \Nch$ to $P(\Nch / \avg{\Nch})$. Red bars are statistical uncertainties, bands are systematic theoretical uncertainties due to sampling from the Bayesian posterior. Black bars are statistical uncertainties, which for \po are smaller than the marker size, and shaded gray boxes are systematic uncertainties.}
  \label{fig:pp_pO_pPb_NchPnch}
\end{figure*}

The difference between the experimentally reported $\Ncoll$ and that produced by IP-Glasma
likely reflects differences in
the Glauber model implementations.
For instance,
for the Woods-Saxon systems IP-Glasma enforces a minimum inter-nucleon distance $d_{\text{min}} = 0.9~\mathrm{fm}$,
larger than the $d_{\text{min}} = 0.4 ~\mathrm{fm}$ typically enforced by experimental Glauber models \cite{Alver:2008aq}.
IP-Glasma also generates multiplicity fluctuations at fixed $N_{\text{part}}$ via subnucleonic fluctuations,
rather than the NBD ansatz used by experiments~\cite{Alver:2008aq,ALICE:2014xsp,ALICE:2018ekf}.
More broadly,
the NBD-plus-participant-scaling ansatz
is conceptually distinct from the IP-Glasma approach,
in which the multiplicity emerges
from the underlying initial-state physics.

\section{Validation of soft and semi-hard particle production}
\label{sec:validation}

To study correlations between hard- and soft-particle production in hadronic collisions,
IP-Glasma must simultaneously provide a realistic description of event activity and high-$p_T$ parton production. 
The former determines how events are classified into centrality classes,
while the latter determines the hard probe multiplicity within those classes. 
We therefore begin by validating the model against charged-particle multiplicity distributions,
and then turn to the high-$p_T$ spectrum.

\subsection{Multiplicity distributions}
\label{sec:multiplicity_distributions}

\begin{figure*}[!t]
	\centering
	\includegraphics[width=0.7\linewidth]{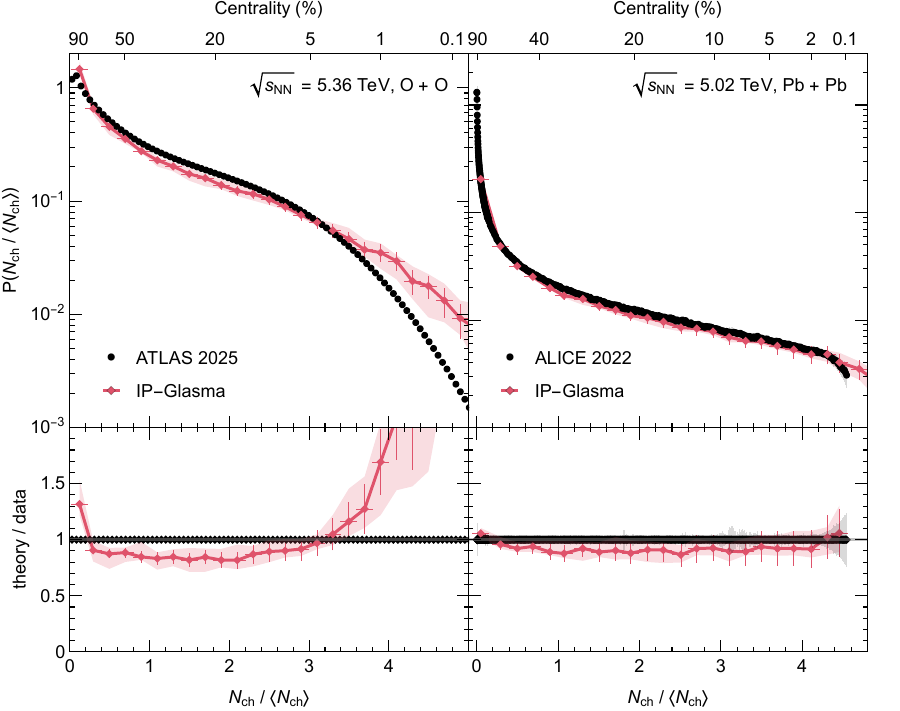}
  \caption{The self-normalized multiplicity distribution $P(\Nch / \avg{\Nch})$ for charged hadrons in $\sqrt{s_{NN}} = 5.36~\mathrm{TeV}$ \oo (left) and $\sqrt{s_{NN}} = 5.02~\mathrm{TeV}$ \pbpb (right) collisions, in the same format as \cref{fig:pp_pO_pPb_NchPnch}. IP-Glasma results are shown in red and experimental data in black: ATLAS~\cite{ATLAS:2025nnt} for \oo, and ALICE~\cite{ALICE:2022xip} for \pbpb. Red bars are statistical uncertainties, bands are systematic theoretical uncertainties due to sampling from the Bayesian posterior. For \pbpb, black bars are statistical uncertainties and shaded gray boxes are systematic uncertainties; no experimental uncertainties were reported in the original ATLAS \oo data~\cite{ATLAS:2025nnt}.}
	\label{fig:OO_PbPb_NchPnch}
\end{figure*}

To assess the ability of IP-Glasma to describe the event activity,
we plot the self-normalized distribution of charged particles $P\left(\Nch / \avg{\Nch} \right)$ as a function of $\Nch / \avg{\Nch}$,
where $\Nch$ is the total number of charged particles produced in an event
and $\avg{\Nch}$ is the mean number of charged particles produced in minimum-bias events;
see \cref{sec:observable_implementation} for a detailed description of how these observables are computed. 
The IP-Glasma results are binned in $\Nch / \avg{\Nch}$ and the center of the bin is chosen according to the prescription in \cite{Lafferty:1994cj}, 
which is important when comparing to data in \pbpb collisions, where the results vary rapidly at low $\Nch / \avg{\Nch}$.
\Cref{fig:pp_pO_pPb_NchPnch} presents the proton-proton and proton-nucleus systems,
\pp, \po, and \ppb,
and \cref{fig:OO_PbPb_NchPnch} the nucleus-nucleus systems,
\oo and \pbpb.
In each case the upper panel shows $P\left(\Nch / \avg{\Nch} \right)$
and the lower panel the ratio of theory to data.

\Cref{fig:pp_pO_pPb_NchPnch} shows that IP-Glasma describes the
charged-particle multiplicity distribution in \pp collisions to within
$20\text{--}30\%$ over the $1\text{--}100\%$ centrality range (left),
with the agreement improving for \po (center) and \ppb (right).
In all systems, however, IP-Glasma overpredicts the yield of the most
central ($\lesssim 1\%$) events, producing too many events with
$\Nch / \avg{\Nch} \gtrsim 4$.
For reference, the $1\%$ centrality cut corresponds to
$\Nch / \avg{\Nch} \simeq 4.1$, $4.1$, $3.3$, $3.9$, and $4.2$
in \pp, \po, \ppb, \oo, and \pbpb collisions, respectively,
as indicated on the upper axes of \cref{fig:pp_pO_pPb_NchPnch,fig:OO_PbPb_NchPnch}.
The origin of this excess is not clear;
the excess may reflect excessive fluctuations of the proton substructure,
for example, in the heavy tail of the log-normal $Q_s^2$
fluctuations from \cref{eqn:saturation_scale_fluctuations}.
Since the discrepancy is confined to centralities below $1\%$,
the excess should not significantly impact our subsequent results,
for which the most central class considered is $0\text{--}5\%$.

Turning to the nucleus-nucleus systems of \cref{fig:OO_PbPb_NchPnch},
the \pbpb description in the right panels agrees remarkably well over the $0.1\text{--}100\%$ centrality range. 
IP-Glasma agrees well with the \oo data in the left panels over the $1\text{--}100\%$ centrality range;
however, the data display a much sharper decrease
at high $\Nch / \avg{\Nch}$
compared to IP-Glasma.

The improved agreement in \po, \ppb, \oo, and \pbpb collisions
compared to \pp collisions
is perhaps unsurprising given that IP-Glasma is valid in the dense--dense regime.
Alternatively,
better agreement may simply be due to the significantly larger role that geometry plays in multiplicity production in the larger systems compared to \pp, 
meaning that high-event-activity \ppb collisions are not as sensitive to fluctuations as high-event-activity \pp collisions.

The high-multiplicity regime is challenging for current hadronic-interaction models generally: the
spread among model predictions grows from ${\sim}20\%$ near the mean multiplicity
to more than an order of magnitude in the tail, with Angantyr%
~\cite{Bierlich:2016smv,Bierlich:2018xfw} and QGSJET~III%
~\cite{Ostapchenko:2024myl,Ostapchenko:2024jsg} under- and overestimating the \po
data, respectively, by an order of magnitude at high $\Nch$, although
DPMJET~III~\cite{Roesler:2000he,Bopp:2005cr} and
EPOS~LHC-R~\cite{Pierog:2025ixr} remain within $30\%$ in this high-multiplicity region~\cite{ATLAS:2026xcm}.
For $N_{\text{ch}} / \left\langle N_{\text{ch}} \right\rangle < 4$, IP-Glasma performs comparably to or better
than the aforementioned state-of-the-art models,
which deviate from data by $25\text{--}50\%$~\cite{ALICE:2022xip,ATLAS:2026xcm}.

It is worth asking to what extent final-state effects,
which are not included in this work but are important in at least \oo and \pbpb collisions,
affect our conclusions.
Entropy is conserved by ideal hydrodynamics,
and so the final multiplicity is not affected in that approximation.
Viscous corrections allow for entropy production,
but the effect is only $\mathcal{O}(10\%)$,
and changes only by $\mathcal{O}(5\%)$ as a function of centrality \cite{Song:2008si}.
Energy loss of high-$p_T$ particles is another important final-state effect, especially in \pbpb collisions.
Because of the power-law spectrum,
we expect any change to the multiplicity caused by final-state energy loss---%
for example extra particles from radiative splittings---%
to have a similarly negligible effect on the $P(\Nch / \avg{\Nch})$ distributions.

\begin{figure}[!t]
	\centering
	\includegraphics[width=\linewidth]{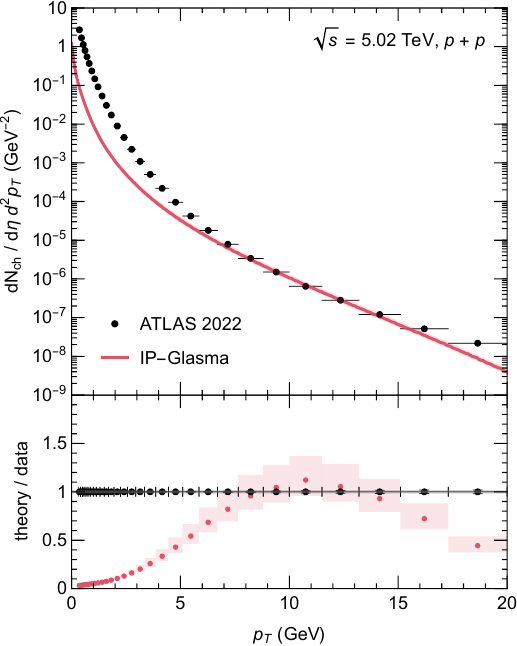}
    \caption{Top: the charged hadron spectrum $d^3 \Nch / d^2 p_T\, d\eta$ in $\sqrt{s} = 5.02~\mathrm{TeV}$ \pp collisions, as a function of transverse momentum $p_T$. IP-Glasma results are shown in red and experimental data from ATLAS~\cite{ATLAS:2022kqu} are shown in black. The ATLAS measurement is reported as an inelastic cross section and has been divided by $\sigma_{\text{inel}}^{NN} = 67.3 \pm 1.2~\mathrm{mb}$~\cite{dEnterria:2020dwq} to convert the cross section to a yield, with the uncertainty on $\sigma_{\text{inel}}^{NN}$ propagated into the experimental systematic uncertainty, where the contribution is negligible. Bottom: ratio of theory to experimental data, with IP-Glasma results binned to match the experimental binning. Red bars are statistical uncertainties, bands are systematic theoretical uncertainties due to sampling from the Bayesian posterior. Black bars are statistical uncertainties and shaded gray boxes are systematic uncertainties.}
	\label{fig:pp_spectrum}
\end{figure}

\subsection{Charged hadron spectra}
\label{sec:charged_hadron_spectra}

Now that we have shown that IP-Glasma can accurately describe the distribution of event activity
across a large variety of systems and centralities,
we turn to the semi-hard physics of particles with $5 ~\mathrm{GeV} \lesssim p_T \lesssim 20 ~\mathrm{GeV}$.
This $p_T$ range corresponds to a typical Bjorken $x$ range of $x \sim p_T / \sqrt{s_{NN}} \sim 0.001\text{--}0.004$,
where the small-$x$ framework is still expected to apply \cite{Gelis:2010nm}. 
\Cref{fig:pp_spectrum} plots the charged hadron spectrum $d^3 \Nch / d^2 p_T\, d\eta$ in $\sqrt{s} = 5.02~\mathrm{TeV}$ \pp collisions as a function of transverse momentum $p_T$.
Results from IP-Glasma are shown as a red curve and experimental data from ATLAS~\cite{ATLAS:2022kqu} are shown as black points.
ATLAS reports this measurement as a charged hadron production cross section, $d^3 \sigma_{\text{ch}} / d^2 p_T\, d\eta$;
we convert $d^3 \sigma_{\text{ch}} / d^2 p_T\, d\eta$ to a yield by dividing by the inelastic nucleon-nucleon cross section
$\sigma_{\text{inel}}^{NN} = 67.3 \pm 1.2 ~\mathrm{mb}$~\cite{dEnterria:2020dwq},
propagating the uncertainty on $\sigma_{\text{inel}}^{NN}$ into the (negligible) experimental systematic uncertainty.
We compare yields rather than cross sections deliberately.
As we will show in \cref{sec:nuclear_modification_factor},
the inelastic cross section predicted by IP-Glasma varies substantially across the Bayesian posterior;
comparing cross sections would therefore fold that poorly constrained normalization
into what is intended to be a comparison of the shape of the spectrum.

\begin{figure}[!t]
	\centering
	\includegraphics[width=\linewidth]{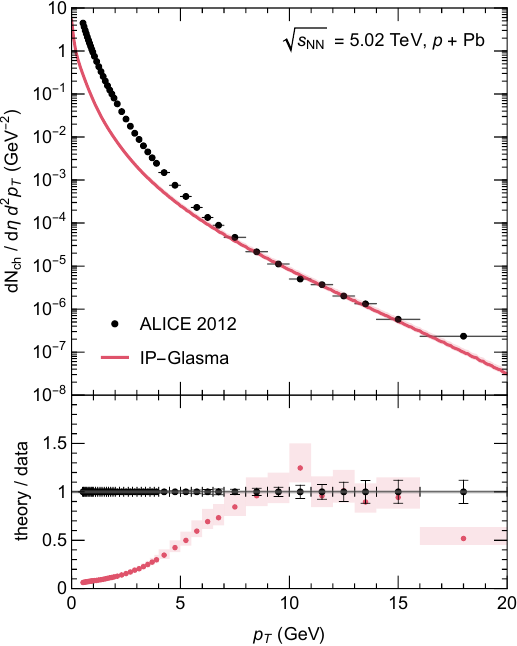}
	\caption{Top: the charged hadron spectrum $d^3 \Nch / d^2 p_T\, d\eta$ in $\sqrt{s_{NN}} = 5.02~\mathrm{TeV}$ \ppb collisions, as a function of transverse momentum $p_T$. IP-Glasma results are shown in red and experimental data from ALICE~\cite{ALICE:2012mj} are shown in black. Bottom: ratio of theory to experimental data, with IP-Glasma results binned to match the experimental binning. Red bars are statistical uncertainties, bands are systematic theoretical uncertainties due to sampling from the Bayesian posterior. Black bars are statistical uncertainties and shaded gray boxes are systematic uncertainties.}
	\label{fig:ppb_spectrum}
\end{figure}

We fix the normalization of the IP-Glasma results to the
$5~\mathrm{GeV} \lesssim p_T \lesssim 15~\mathrm{GeV}$ spectrum
in $\sqrt{s} = 5.02~\mathrm{TeV}$ \pp collisions; see \cref{fig:pp_spectrum}.
The same normalization is used for every system and collision energy considered in this work.
The normalization does not carry a physical interpretation; the MV model~\cite{McLerran:1993ka,McLerran:1994vd} produces a spectrum considerably harder than the data,
so the normalization needed to bring the two into agreement
depends on the range in $p_T$ over which the matching is performed.
We therefore interpret the need for a normalization as a deficiency in the MV model,
as opposed to, for example, higher-order effects.
Prior work has found that a better description of both
diffractive meson data and the $p_T$ spectrum in hadronic collisions
can be obtained by introducing an anomalous dimension $\gamma$ to the MV model \cite{Dumitru:2005gt, ALbacete:2010ad, Tribedy:2010ab,Lappi:2013zma}; 
we leave an implementation of the anomalous dimension $\gamma$ in IP-Glasma for future work.

\Cref{fig:pp_spectrum,fig:ppb_spectrum} establish
that the slope of the spectrum over $5~\mathrm{GeV} \lesssim p_T \lesssim 15~\mathrm{GeV}$
is approximately reproduced
in \pp collisions and, with no further adjustment,
in $\sqrt{s_{NN}} = 5.02~\mathrm{TeV}$ \ppb collisions.
The slope is the only feature of the spectrum on which our results depend.
The slope in this range is, however, sensitive to the lattice spacing $a$:
because the fragmentation convolution in \cref{eqn:hadron_spectrum}
samples gluons at $p_T^g = p_T^h / z > p_T^h$,
the integral over $z$ is truncated by the ultraviolet cutoff of the lattice,
and the hadron spectrum is depleted at momenta well below that cutoff.
We quantify this effect in \cref{sec:dependence_of_results_on_lattice_spacing},
and show that while the spectra are very sensitive to the lattice spacing, 
this sensitivity largely cancels in observables composed of ratios,
leaving a residual ${\lesssim}20\%$ uncertainty in the normalization of $R_{AB}$.

\section{Minimum-bias nuclear modification factor}
\label{sec:nuclear_modification_factor}

Before discussing the impact of hard-soft correlations on the centrality-dependent nuclear modification factor,
we assess IP-Glasma's ability to describe the minimum-bias case.
The nuclear modification factor $R_{AB}$ in the collision $A + B$ 
is computed according to \cref{eqn:nuclear_modification_factor},
as described in detail in \cref{sec:observable_implementation}.
Importantly, 
the number of binary collisions $\Ncoll$,
which appears in the denominator of the $R_{AB}$ ratio,
is not computed in our theoretical framework but rather taken from values quoted by the relevant experiment.
Using the experimental value for $\Ncoll$ is a deliberate choice:
the choice means that our results do not depend on the details or applicability of the particular Glauber model
that we would otherwise have to adopt ourselves; see \cref{sec:observable_implementation} for a more detailed discussion of the difference between the experimentally reported \Ncoll and that computed in IP-Glasma.

\begin{figure}[!t]
	\centering
	\includegraphics[width=\linewidth]{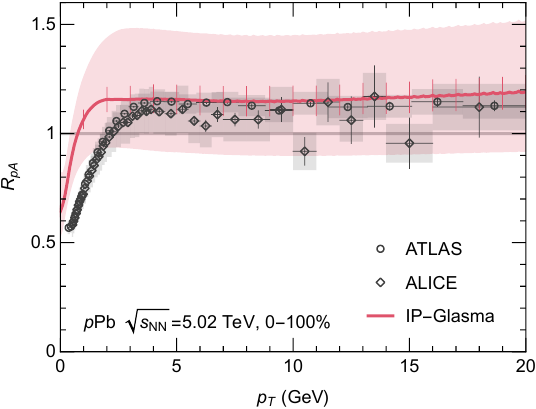}
	\caption{The minimum-bias nuclear modification factor $R_{pA}$ of \cref{eqn:nuclear_modification_factor}, normalized by the experimentally reported \Ncoll, for charged hadrons produced in $0\text{--}100\%$ centrality \ppb collisions at $\sqrt{s_{NN}} = 5.02~\mathrm{TeV}$. IP-Glasma results are shown in red and experimental data from ALICE~\cite{ALICE:2014xsp} and ATLAS \cite{ATLAS:2022kqu} are shown in black. Red bars are statistical uncertainties, bands are systematic theoretical uncertainties due to sampling from the Bayesian posterior. Black bars are statistical uncertainties and shaded gray boxes are systematic experimental uncertainties.}
	\label{fig:RAA_pPb_MB}
\end{figure}

\Cref{fig:RAA_pPb_MB} plots the \Ncoll-normalized nuclear modification factor $R_{pA}$
of \cref{eqn:nuclear_modification_factor}
as a function of $p_T$ for charged hadrons
produced in $0\text{--}100\%$ $\sqrt{s_{NN}} = 5.02 ~\mathrm{TeV}$ \ppb collisions.
We observe
that IP-Glasma predicts $R_{pA} \simeq 1.2 \pm 0.3$
for $p_T \gtrsim 5 ~\mathrm{GeV}$,
in good agreement with experimental data $R_{pA} \simeq 1.1 \pm 0.1$.
Considering the $p_T \lesssim 5 ~\mathrm{GeV}$ results in \cref{fig:RAA_pPb_MB}, 
we see that the IP-Glasma result is less suppressed than the data.
The disagreement with data is perhaps unsurprising 
given that the use of fragmentation functions is not valid at such low $p_T$. 
In \cref{sec:dependence_of_results_on_physical_choices} we show that our results are insensitive to the choice of fragmentation function.
If we were to implement more realistic hadronization that accounts for the possibility that partons at low $p_T$ do not always lose energy when forming hadrons,
then we anticipate better agreement with data~\cite{Greif:2020rhi}.
The shape of $R_{pA}$ in the low-$p_T$ region is, however, sensitive to the scale at which the coupling runs.
\Cref{sec:dependence_of_results_on_physical_choices} shows that evaluating the coupling at
$\tilde{\mu} = Q_s^{\text{max}}$ rather than at our default $\tilde{\mu} = p_T/2$
produces a pronounced Cronin-like peak, in significant tension with data.
Evaluating the simulation at fixed coupling does not significantly affect the minimum-bias $R_{pA}$.

Two features of the $p_T \gtrsim 5 ~\mathrm{GeV}$ prediction stand out.
The first is the width of the systematic band%
\footnote{
To remind the reader,
the systematic band is formed as the $16$th--$84$th percentile range of the 20 posterior samples sampled from the Bayesian analysis on HERA data; see \cref{sec:bayesian_posterior_sampling}.
},
${\pm}\,0.3$, which is three times the experimental uncertainty
and far larger than the statistical uncertainty of the calculation.
The second is that many of the posteriors
do not converge to $R_{pA} \sim 1$ at high $p_T$,
whereas such convergence is typically expected from analytic considerations in the CGC \cite{Kharzeev:2002pc,Kharzeev:2003wz,Lappi:2013zma}.
We show in the remainder of this section that both features
are statements about the inelastic nucleon-nucleon cross section $\sigma_{\text{inel}}^{NN}$
rather than about hard particle production,
and that the parameter sets which reproduce the measured $\sigma_{\text{inel}}^{NN}$
are precisely those which reproduce the measured $R_{pA}$.

\begin{figure}[!t]
	\centering
	\includegraphics[width=\linewidth]{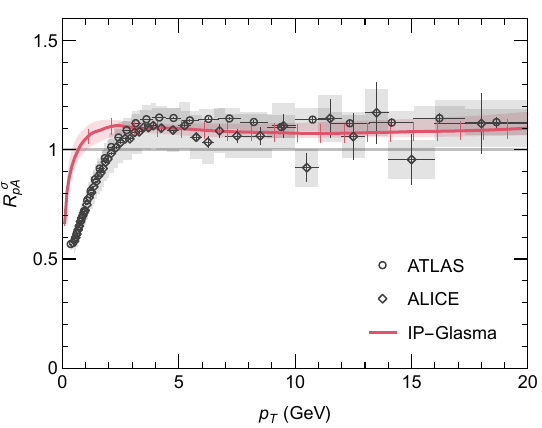}
	\caption{The cross-section-normalized minimum-bias nuclear modification factor $R^{\sigma}_{pA}$ of \cref{eqn:nuclear_modification_factor_sigma}, which makes no reference to \Ncoll at all, for charged hadrons produced in $0\text{--}100\%$ centrality \ppb collisions at $\sqrt{s_{NN}} = 5.02~\mathrm{TeV}$. IP-Glasma results are shown in red and experimental data for $R_{pA}$ from \cref{eqn:nuclear_modification_factor} from ALICE~\cite{ALICE:2014xsp} and ATLAS \cite{ATLAS:2022kqu} are shown in black. Red bars are statistical uncertainties, bands are systematic theoretical uncertainties due to sampling from the Bayesian posterior. Black bars are statistical uncertainties and shaded gray boxes are systematic experimental uncertainties. These are the same simulated events as \cref{fig:RAA_pPb_MB} and differ from that figure only in their normalization.}
	\label{fig:RAA_pPb_MB_sigma}
\end{figure}

The quantity $R_{AB}^{\sigma}$ defined in \cref{eqn:nuclear_modification_factor_sigma} does not rely on the Glauber model for its normalization
and,
therefore,
one might anticipate that $R_{AB}^{\sigma}$ is better behaved asymptotically.
\Cref{fig:RAA_pPb_MB_sigma} plots the minimum-bias $R_{pA}^{\sigma}$ in $\sqrt{s_{NN}} = 5.02 ~\mathrm{TeV}$ \ppb collisions as a function of $p_T$ compared to the same experimental data as \cref{fig:RAA_pPb_MB}. 
We observe in the figure
that IP-Glasma predicts $R_{pA}^{\sigma} = 1.1 \pm 0.05$, in very good agreement with data.
Of particular note
is that the uncertainties of $R_{pA}^{\sigma}$ are a factor of six smaller than those of $R_{pA}$, and that $R_{pA}^{\sigma}$ is not very different from one for all posteriors.
The 10\% enhancement in $R_{pA}^{\sigma}$ observed in \cref{fig:RAA_pPb_MB_sigma} may arise from residual lattice effects, which we discuss in \cref{sec:dependence_of_results_on_lattice_spacing}, or may have a physical origin.
Disentangling these possibilities would require running the entire Bayesian posterior at a finer lattice spacing, which is computationally expensive because the computation time scales as the inverse cube of the lattice spacing.
We leave such an investigation for future work.

The stark difference in the size of the uncertainties between $R_{pA}$ and $R_{pA}^{\sigma}$ 
may be understood through the following approximate relation,
which we derive in \cref{sec:inelastic_cross_section_parameters}:
\begin{equation}
  R_{AB} \simeq R^{\sigma}_{AB} \; \frac{\sigma_{\text{inel}}^{pp,\,\text{IPG}}}{\sigma_{\text{inel}}^{NN,\,\text{Glauber}}},
  \label{eqn:rab_normalization}
\end{equation}
where $R^{\sigma}_{AB}$ is from \cref{eqn:nuclear_modification_factor_sigma},
$\sigma_{\text{inel}}^{pp,\,\text{IPG}}$ is the \pp inelastic cross section that IP-Glasma produces,
and $\sigma_{\text{inel}}^{NN,\,\text{Glauber}}$ is the inelastic nucleon-nucleon cross section that was used in the Glauber model computation of $N_{\text{coll}}$ by ALICE~\cite{ALICE:2014xsp}.
\Cref{eqn:rab_normalization} implies that the observed difference between $R_{pA}$ and $R_{pA}^{\sigma}$
is due to a difference in the inelastic nucleon-nucleon cross section predicted by IP-Glasma and that used in the determination of $N_{\text{coll}}$ with the Glauber model.
Interestingly,
several CGC treatments also find $R_{pA} \neq 1$ at high $p_T$ \cite{Albacete:2010bs,Albacete:2012xq,Tribedy:2011aa},
which is perhaps an indication that those calculations suffer from the same normalization issue.

To illustrate \cref{eqn:rab_normalization},
we compute the inelastic proton-proton cross section $\sigma_{\text{inel}}^{pp}$
according to \cref{eqn:inelastic_cross_section} using IP-Glasma \pp simulations,
and assume that $\sigma_{\text{inel}}^{pp} \simeq \sigma_{\text{inel}}^{NN}$.
We find $\sigma_{\text{inel}}^{pp} = 74 \pm 25 \text{ (sys.)} \pm 1 \text{ (stat.)} ~\mathrm{mb}$,
implying that a significant fraction of the posteriors are in tension with the
$\sigma_{\text{inel}}^{NN} = 70 \pm 5 ~\mathrm{mb}$
that enters the Glauber determination of the \Ncoll used to normalize $R_{pA}$ in \cref{fig:RAA_pPb_MB} \cite{ALICE:2014xsp}.\footnote{%
Inelastic nucleon-nucleon cross sections enter this work in three distinct roles,
of which only the last is a prediction of our framework.
$70 \pm 5 ~\mathrm{mb}$ is the value entering the Glauber model from which ALICE extracts the reported \Ncoll
at $\sqrt{s_{NN}} = 5.02 ~\mathrm{TeV}$
\cite{ALICE:2014xsp}, which $R_{AB}$ presented here therefore inherits.
$67 ~\mathrm{mb}$ at $\sqrt{s_{NN}} = 5.02 ~\mathrm{TeV}$,
$70 ~\mathrm{mb}$ at $5.36 ~\mathrm{TeV}$,
and $74.6 ~\mathrm{mb}$ at $9.62 ~\mathrm{TeV}$ \cite{dEnterria:2020dwq}
are the inputs to the Monte Carlo Glauber model
that we run alongside IP-Glasma to obtain the \Ncoll of \cref{fig:ncoll_combined_vs_centrality}
and of \cref{fig:RAA_system_comparison};
because that model takes $\sigma_{\text{inel}}^{NN}$ as an input rather than predicting it,
the mismatch discussed here is not visible in \cref{fig:ncoll_combined_vs_centrality}.
$74 \pm 25 ~\mathrm{mb}$ is the $\sigma_{\text{inel}}^{pp}$ that IP-Glasma predicts.}

\begin{figure}[!t]
	\centering
	\includegraphics[width=\linewidth]{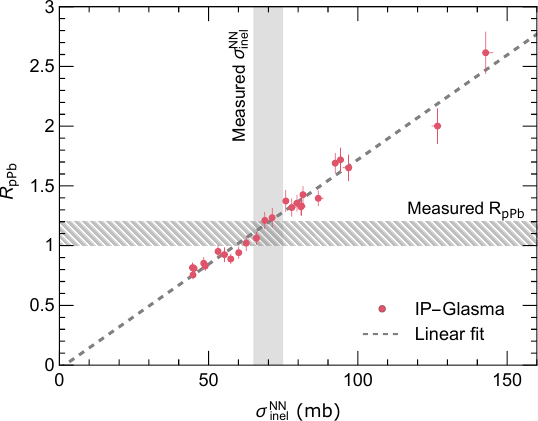}
\caption{The minimum-bias nuclear modification factor $R_{p\pb}$ for $p_T \simeq 8~\mathrm{GeV}$ charged hadrons in $\sqrt{s_{NN}} = 5.02~\mathrm{TeV}$ \ppb collisions, plotted against the inelastic proton-proton cross section $\sigma_{\text{inel}}^{pp}$ that IP-Glasma predicts for the same parameter set, with one point per posterior sample (red). Bars are statistical uncertainties and the dashed gray line is a linear fit. The vertical gray band is the measured $\sigma_{\text{inel}}^{NN} = 70 \pm 5~\mathrm{mb}$~\cite{ALICE:2014xsp}, which enters the Glauber determination of \Ncoll; the horizontal hatched gray band is the measured $R_{pA}$ at the same $p_T$~\cite{ALICE:2014xsp,ATLAS:2022kqu}.}
	\label{fig:cross_section_vs_rpa}
\end{figure}

\Cref{fig:cross_section_vs_rpa} plots, for each posterior sample,
the $R_{pA}$ that IP-Glasma predicts for $p_T \simeq 8~\mathrm{GeV}$ charged hadrons
in minimum-bias $\sqrt{s_{NN}} = 5.02~\mathrm{TeV}$ \ppb collisions
against the $\sigma_{\text{inel}}^{pp}$ predicted by that same sample.
We also show, by gray bands, the experimentally measured values $\pm 1 \sigma$ of $R_{p\mathrm{Pb}}$ and $\sigma_{\text{inel}}^{NN}$.
We see in \cref{fig:cross_section_vs_rpa} that
$R_{pA}$ and $\sigma_{\text{inel}}^{pp}$ are linearly correlated
with a fit that passes close to the origin,
consistent with \cref{eqn:rab_normalization}.
Notice that the IP-Glasma cross sections span roughly $45\text{--}145 ~\mathrm{mb}$
and most of the posterior is thus in significant tension with 
the value of $\sigma_{\text{inel}}^{NN}$ used in the Glauber determination of \Ncoll.

\begin{figure}[!t]
	\centering
	\includegraphics[width=\linewidth]{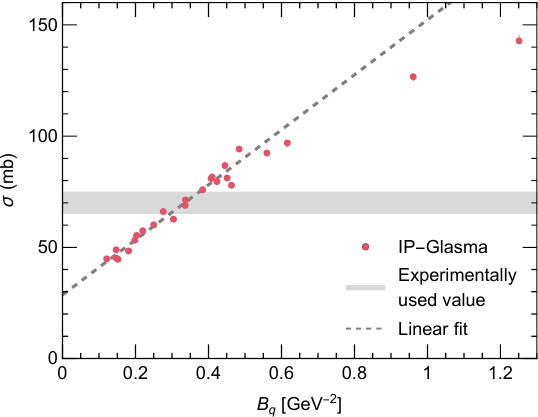}
\caption{The inelastic proton-proton cross section $\sigma_{\text{inel}}^{pp}$ at $\sqrt{s} = 5.02~\mathrm{TeV}$, calculated from 20 IP-Glasma parameter sets sampled from the HERA Bayesian posterior (red). Results are shown for each posterior sample as a function of the corresponding hotspot size parameter $B_q$. The gray band indicates the value used in the Glauber determination of \Ncoll, $\sigma_{\text{inel}}^{NN} = 70 \pm 5~\mathrm{mb}$~\cite{ALICE:2014xsp}. A linear fit to the posteriors with $B_q < 0.5 ~\mathrm{GeV}^{-2}$ is displayed as a dashed gray line.}
	\label{fig:cross_section_vs_bgq}
\end{figure}

\Cref{fig:cross_section_vs_rpa} also shows the consistency of the prediction with data:
the vertical band marking the measured $\sigma_{\text{inel}}^{NN}$
and the horizontal band marking the measured $R_{pA}$
intersect where the posterior samples pass through.
The parameter sets that reproduce the measured inelastic cross section
are therefore precisely those that reproduce the measured $R_{pA}$,
and constraining the posterior on $\sigma_{\text{inel}}^{pp}$ would not merely shrink
the theoretical uncertainty of \cref{fig:RAA_pPb_MB} but collapse the IP-Glasma prediction onto the data.
We find that the spread in $\sigma_{\text{inel}}^{pp}$ across the posterior
is dominated by the hotspot size parameter $B_q$.
\Cref{fig:cross_section_vs_bgq} plots $\sigma_{\text{inel}}^{pp}$ for each of the 20 posterior samples
against the corresponding $B_q$,
compared to the $\sigma_{\text{inel}}^{NN} = 70 \pm 5 ~\mathrm{mb}$ used in the Glauber determination of \Ncoll.
The dependence is linear for $B_q \lesssim 1 ~\mathrm{GeV}^{-2}$,
as increasing the area of the hotspots linearly increases the cross section.
For $B_q \gtrsim 1~\mathrm{GeV}^{-2}$,
we observe deviations from linearity,
which we attribute to overlap between the hotspots,
as $N_q B_q \sim 3~\mathrm{GeV}^{-2}$ becomes comparable to $B_{qc} \sim 4.45~\mathrm{GeV}^{-2}$, the characteristic size of the proton.

The strong sensitivity of $\sigma_{\text{inel}}^{pp}$ to $B_q$ suggests that the measured inelastic \pp cross section can provide a direct constraint on $B_q$.
However,
we cannot yet infer from \cref{fig:cross_section_vs_bgq}
that $B_q$ must be ${\sim} 0.3\text{--}0.5 ~\mathrm{GeV}^{-2}$,
because of an uncertainty in how we define an inelastic collision.
As discussed in \cref{sec:observable_implementation}, 
we classify a collision as inelastic 
if the thickness functions of both nuclei exceed a threshold,
$T_A > T_{\min}$ and $T_B > T_{\min}$,
in at least one lattice cell. 
In all results shown in this work, $T_{\min} = 10^{-4}~\mathrm{GeV}^2$, corresponding to $Q_s \simeq 0.04~\mathrm{GeV}$ at $y=0$. 
Raising the threshold to the perhaps more physical $Q_s = 0.2~\mathrm{GeV}$,
comparable to $\Lambda_{\text{QCD}}$ and to the infrared regulator $m$,
reduces $\sigma_{\text{inel}}^{pp}$ from $73$ to $51~\mathrm{mb}$ at the MAP parameter set.
Therefore,
there is a degeneracy between the parameters that determine $\sigma_{\text{inel}}^{NN}$,
and so the measured $\sigma_{\text{inel}}^{NN}$ cannot on its own be used to constrain $B_q$.
A dedicated study, using $\sigma_{\text{inel}}^{NN}$ together with its $\sqrt{s}$ dependence and perhaps other observables, like the tail of the multiplicity distributions, could constrain both parameters. Such a study would also have to address how diffractive events---included in the measured $\sigma_{\text{inel}}$ but absent from our framework---should be treated, and whether a cut on $T_A$, on $Q_s$, or on the produced energy density is the most appropriate definition of an inelastic collision.

\begin{figure}[!t]
  \centering
  \includegraphics[width=\linewidth]{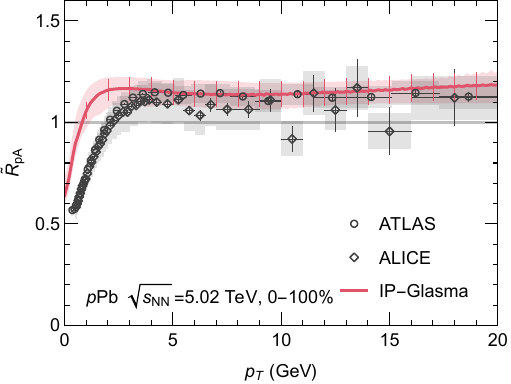}
  \caption{The rescaled minimum-bias nuclear modification factor $\tilde{R}_{pA}$ of \cref{eqn:rescaled_raa} for charged hadrons produced in $0\text{--}100\%$ centrality \ppb collisions at $\sqrt{s_{NN}} = 5.02~\mathrm{TeV}$, obtained by dividing out of $R_{pA}$, for each posterior sample, the ratio of the \pp cross section that IP-Glasma predicts for that sample to the inelastic nucleon-nucleon cross section used in the Glauber determination of \Ncoll. These are the same simulated events as \cref{fig:RAA_pPb_MB} and differ from that figure only in their normalization; \cref{fig:RAA_pPb_MB_sigma} removes the same mismatch in an alternative way, through the cross-section-normalized $R^{\sigma}_{pA}$ computed directly rather than by rescaling $R_{pA}$. IP-Glasma results are shown in red and experimental data from ALICE~\cite{ALICE:2014xsp} and ATLAS \cite{ATLAS:2022kqu} are shown in black. Red bars are statistical uncertainties and bands are systematic theoretical uncertainties due to sampling from the Bayesian posterior. Black bars are statistical uncertainties and shaded gray boxes are systematic experimental uncertainties.}
  \label{fig:RAA_pPb_MB_rescaled}
\end{figure}

Before we can discuss the centrality-dependent $R_{AB}$,
it would be helpful to,
at least partially,
resolve this normalization issue in $R_{AB}$
in a way that can be applied to the centrality-cut case.
We saw in \cref{fig:RAA_pPb_MB_sigma} that $R_{pA}^{\sigma}$ removes the large uncertainty in the normalization of the nuclear modification factor,
which we understood from \cref{eqn:rab_normalization}. 
However,
$R_{pA}^{\sigma}$ is only defined for minimum-bias collisions.
In principle,
to reduce the normalization uncertainty of $R_{pA}$ from \cref{eqn:nuclear_modification_factor},
which is defined for centrality-cut collisions,
one should constrain both $B_q$ and $T_{\text{min}}$ on the measured $\sigma_{\text{inel}}$.
However,
as discussed in the preceding paragraph, such a constraint would take substantial extra effort.
In lieu of these constraints,
we instead use \cref{eqn:rab_normalization} to construct a ``rescaled'' $R_{pA}$
\begin{equation}
  \tilde{R}_{pA} \equiv R_{pA} \, \frac{\sigma_{\text{inel}}^{NN,\,\text{Glauber}}}{\sigma_{\text{inel}}^{pp,\,\text{IPG}}} \, ,
  \label{eqn:rescaled_raa}
\end{equation}
using the $R_{pA}$ and the $\sigma_{\text{inel}}^{pp,\,\text{IPG}}$ of \cref{eqn:inelastic_cross_section}
predicted by each posterior sample,
and $\sigma_{\text{inel}}^{NN,\,\text{Glauber}}$ given by the inelastic nucleon-nucleon cross section used in the Glauber determination of the \Ncoll that normalizes $R_{pA}$.
\Cref{fig:RAA_pPb_MB_rescaled} shows $\tilde{R}_{pA}$ from \cref{eqn:rescaled_raa} for minimum-bias $\sqrt{s_{NN}} = 5.02 ~\mathrm{TeV}$ \ppb collisions as a function of $p_T$ with the same data as \cref{fig:RAA_pPb_MB}.
We see in the figure a reduction of the uncertainties similar to that of $R_{pA}^{\sigma}$ in \cref{fig:RAA_pPb_MB_sigma} relative to the standard $R_{pA}$.
We will also use \cref{eqn:rescaled_raa} for the centrality-dependent $R_{AA}$ and $R_{pA}$ in \cref{sec:centralitydependent_nuclear_modification_factor},
where the rescaling works well in all but the most peripheral classes.

\begin{figure*}[!t]
  \centering
  \includegraphics[width=0.49\linewidth]{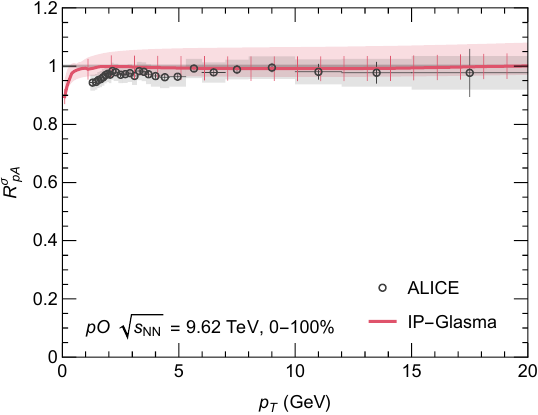}
  \hfill
  \includegraphics[width=0.49\linewidth]{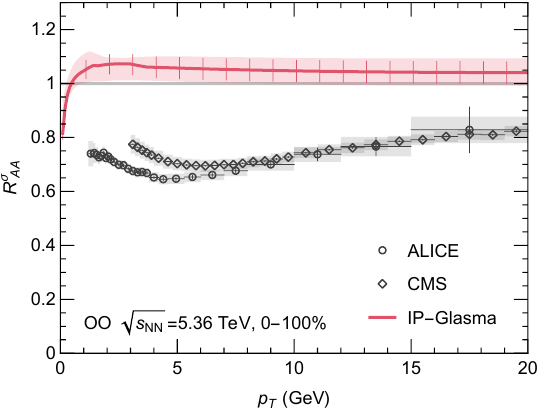}
  \caption{Minimum-bias nuclear modification factors for oxygen collision systems as a function of hadron transverse momentum $p_T$, in the cross-section-normalized form of \cref{eqn:nuclear_modification_factor_sigma}. Left: $R^{\sigma}_{pO}$ for charged hadrons produced in $0\text{--}100\%$ centrality \po collisions at $\sqrt{s_{NN}} = 9.62~\mathrm{TeV}$. Right: $R^{\sigma}_{OO}$ for charged hadrons produced in $0\text{--}100\%$ centrality \oo collisions at $\sqrt{s_{NN}} = 5.36~\mathrm{TeV}$. IP-Glasma results are shown in red and experimental data are shown in black, from ALICE~\cite{ALICE:2026zck} (circles, both panels) and CMS~\cite{CMS:2025bta} (diamonds, right panel only). Red bars are statistical uncertainties, bands are systematic theoretical uncertainties due to sampling from the Bayesian posterior. Black bars are statistical uncertainties and shaded gray boxes are systematic uncertainties.}
  \label{fig:oxygen_MB}
\end{figure*}

\Cref{fig:oxygen_MB} shows the minimum-bias results from the recent run of oxygen ions at the LHC.
Since $R^{\sigma}_{AB}$ is the quantity measured experimentally for the oxygen collisions \cite{ALICE:2026zck,CMS:2025bta},
we compare directly to the data using $R^{\sigma}_{AB}$ rather than either $R_{AB}$ or $\tilde{R}_{AB}$ discussed previously.
In the left panel of \cref{fig:oxygen_MB}, we plot the $R^{\sigma}_{pO}$ from IP-Glasma of charged hadrons produced in $0\text{--}100\%$ centrality \po collisions at $\sqrt{s_{NN}} = 9.62 ~\mathrm{TeV}$ (red line),
together with the measured $R^{\sigma}_{pO}$ of
neutral pions by ALICE~\cite{ALICE:2026zck} (black circles)
as a function of $p_T$.
In the right panel of \cref{fig:oxygen_MB}, we plot the $R^{\sigma}_{OO}$
from IP-Glasma of
charged hadrons produced in $0\text{--}100\%$ centrality \oo collisions at $\sqrt{s_{NN}} = 5.36 ~\mathrm{TeV}$ (red line),
together with the measured $R^{\sigma}_{OO}$ of
charged hadrons by CMS \cite{CMS:2025bta} (black diamonds)
and of neutral pions by ALICE~\cite{ALICE:2026zck} (black circles)
as a function of $p_T$.

From the left and right panels of \cref{fig:oxygen_MB}, respectively,
we see that IP-Glasma predicts 
$R^{\sigma}_{pO} \simeq 1.01 \pm 0.05 ~\text{(sys.)} \pm 0.04 ~\text{(stat.)}$ and
$R^{\sigma}_{OO} \simeq 1.05 \pm 0.05 ~\text{(sys.)} \pm 0.04~\text{(stat.)}$ for $p_T \gtrsim 3 ~\mathrm{GeV}$, consistent with no nuclear modification.
Because the systematic uncertainties are comparable to the statistical uncertainties,
one should view the systematic uncertainty estimate as an upper bound for the true systematic uncertainty; see \cref{sec:bayesian_posterior_sampling}.
Considering the left panel of \cref{fig:oxygen_MB}, we see that IP-Glasma describes both the normalization and the lack of $p_T$ dependence of the \po data extremely well.
The right panel of \cref{fig:oxygen_MB}, on the other hand,
shows that IP-Glasma describes neither the normalization nor the shape of the experimental \oo data, supporting the scenario of final-state partonic energy loss leading to $R^{\sigma}_{OO} < 1$ \cite{Faraday:2025pto, ALICE:2026zck}.

The double ratio $R^{\sigma}_{AA} / {R^{\sigma}_{pA}}^2$ was proposed in \cite{Jonas:2026yoz}
as a way to reduce the sensitivity to the nPDFs, which cancel in the double ratio to within a few percent.
Subsequently,
$R^{\sigma}_{OO} / {R^{\sigma}_{pO}}^2$ has been measured by the ALICE collaboration \cite{ALICE:2026zck}.
We find $R^{\sigma}_{OO} / {R^{\sigma}_{pO}}^2 \simeq \num{1.07(12:04)} ~\text{(sys.)} \pm 0.09 ~\text{(stat.)}$,
consistent with unity and above the measured ${\simeq}\,0.7$ over $5 \lesssim p_T \lesssim 10 ~\mathrm{GeV}$,
again consistent with the interpretation of significant final-state energy loss.
Because we predict $R^{\sigma}_{pO} \simeq 1$,
the double ratio is for us numerically close to $R^{\sigma}_{OO}$ itself
and adds little beyond the right panel of \cref{fig:oxygen_MB};
we show the IP-Glasma prediction for $R^{\sigma}_{OO} / {R^{\sigma}_{pO}}^2$,
and discuss why the double ratio does not reduce our theoretical uncertainty,
in \cref{sec:double_ratio}.

\section{Centrality-dependent nuclear modification factor}
\label{sec:centralitydependent_nuclear_modification_factor}

We now present the main result of this work:
the centrality-dependent nuclear modification factor $R_{AB}$,
in which the hard-soft correlations of IP-Glasma are most directly visible.
We will show that IP-Glasma reproduces the measured centrality dependence of $R_{pA}$ in \ppb collisions,
and qualitatively captures the sharp decrease of $R_{AA}$ toward the most peripheral \pbpb collisions,
despite being constrained only by HERA data.
Before presenting these results,
we discuss two known biases associated with CL1-based centrality determination
that are relevant for interpreting the comparison.

The first is the ``multiplicity bias''~\cite{ALICE:2014xsp}.
Because the CL1 detector ($|\eta| < 1.4$) fully overlaps with the tracking region,
multiplicity-based centrality selection is maximally correlated
with the same nucleon-by-nucleon fluctuations that drive fluctuations 
in hard-particle production 
at equal $\Ncoll$ and $N_{\text{part}}$.
Because IP-Glasma simultaneously describes 
soft- and hard-particle production
with the same underlying framework,
this multiplicity bias is captured by our model,
making the midrapidity centrality selection the most natural for our comparison.

The second is the ``jet-veto bias''~\cite{ALICE:2014xsp}.
In peripheral \ppb collisions,
the total CL1 multiplicity is small, $\mathcal{O}(10)$ charged particles \cite{ALICE:2014xsp},
so the $\mathcal{O}(5)$ particles in a single jet \cite{ALICE:2023oww}
constitute a significant fraction of the event multiplicity.
A single jet can therefore migrate an event from a peripheral into a more central class.
In this work we hadronize with fragmentation functions,
which do not track the additional particles produced by radiative splittings.
The multiplicity associated with a jet is therefore suppressed relative to reality,
and the jet-veto bias is absent from our model.

A final technical remark concerns the normalization of the results presented below.
Centrality selection requires the $\Ncoll$-normalized $R_{AB}$
defined in \cref{eqn:nuclear_modification_factor}.
As shown in \cref{sec:nuclear_modification_factor}, the theoretical uncertainty of $R_{AB}$ in IP-Glasma is dominated by the spread in $\sigma_{\text{inel}}^{pp}$ across the posterior.
The dependence of $R_{AB}$ on $\sigma_{\text{inel}}^{pp}$ enters \cref{eqn:rab_normalization} as an overall multiplicative offset that is fixed for each posterior sample and independent of $p_T$ and relatively independent of centrality.
We therefore remove this offset sample by sample and plot the $\tilde{R}_{AB}$ defined in \cref{eqn:rescaled_raa}.
The central values are essentially unchanged from the $R_{AB}$ of \cref{eqn:nuclear_modification_factor},
while the systematic error band is reduced by roughly a factor of four,
as for the minimum-bias case in \cref{fig:RAA_pPb_MB_rescaled},
except in the most peripheral classes, where the rescaled uncertainties are slightly larger than the unrescaled ones.
We anticipate that $\tilde{R}_{AB}$ will be a good approximation for $R_{AB}$ after constraining the posteriors on the measured inelastic nucleon-nucleon cross section.
The corresponding unrescaled results for $R_{AB}$ are shown in \cref{sec:unrescaled_centrality}.

\begin{figure}[!t]
	\centering
	\includegraphics[width=\linewidth]{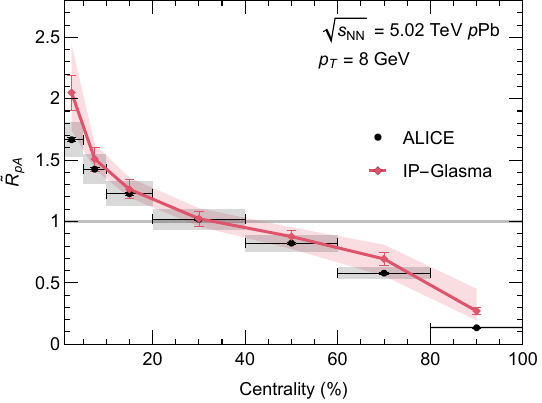}
	\caption{The rescaled nuclear modification factor $\tilde{R}_{pA}$ of \cref{eqn:rescaled_raa} for charged hadrons with $7.5~\mathrm{GeV} \leq p_T \leq 8.5~\mathrm{GeV}$ produced in \ppb collisions at $\sqrt{s_{NN}} = 5.02~\mathrm{TeV}$ as a function of centrality. IP-Glasma results are shown in red and experimental data from ALICE~\cite{ALICE:2014xsp} are shown in black. Red bars are statistical uncertainties, bands are systematic theoretical uncertainties due to sampling from the Bayesian posterior. Black bars are statistical uncertainties and shaded gray boxes are systematic uncertainties. The unrescaled $R_{pA}$ is shown in \cref{fig:RpA_vs_centrality_pPb_502TeV_unrescaled}.}
	\label{fig:RpA_vs_centrality_pPb_502TeV}
\end{figure}

\begin{figure*}[!htbp]
  \centering
  \includegraphics[width=\linewidth]{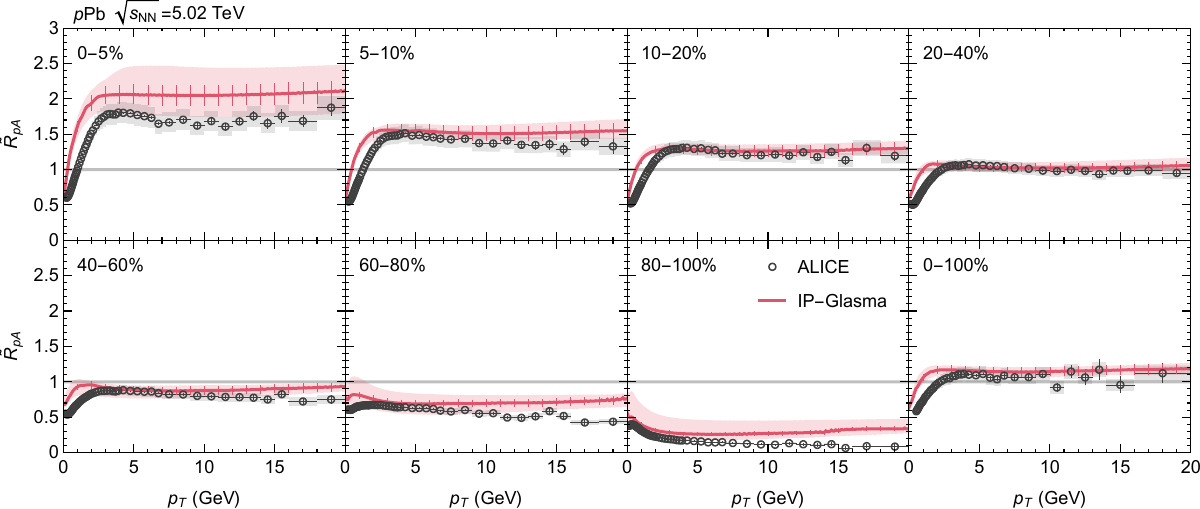}
  \caption{The rescaled nuclear modification factor $\tilde{R}_{pA}$ of \cref{eqn:rescaled_raa} for charged hadrons produced in \ppb collisions at $\sqrt{s_{NN}} = 5.02~\mathrm{TeV}$ as a function of $p_T$ in the $0\text{--}5\%$, $5\text{--}10\%$, $10\text{--}20\%$, $20\text{--}40\%$, $40\text{--}60\%$, $60\text{--}80\%$, and $80\text{--}100\%$ centrality classes. The bottom-right panel shows the minimum-bias ($0\text{--}100\%$) result of \cref{fig:RAA_pPb_MB_rescaled}. IP-Glasma results are shown in red and experimental data from ALICE~\cite{ALICE:2014xsp} are shown in black. Red bars are statistical uncertainties and bands are systematic theoretical uncertainties due to sampling from the Bayesian posterior. Black bars are statistical uncertainties and shaded gray boxes are systematic uncertainties.}
  \label{fig:rppb_corrected_cent_vs_pt}
\end{figure*}

\Cref{fig:RpA_vs_centrality_pPb_502TeV} plots the rescaled nuclear modification factor $\tilde{R}_{pA}$ as a function of centrality
for charged hadrons in $\sqrt{s_{NN}} = 5.02~\mathrm{TeV}$ \ppb collisions,
with IP-Glasma results shown in red
and ALICE data~\cite{ALICE:2014xsp} in black.
We select the $p_T \simeq 8~\mathrm{GeV}$ bin.
The centrality-selected $\tilde{R}_{pA}$ does not vary significantly with $p_T$ for $p_T \gtrsim 8~\mathrm{GeV}$
and, therefore,
our conclusions are not sensitive to the particular $p_T$ bin chosen;
the full $p_T$ dependence in each centrality class is shown in \cref{fig:rppb_corrected_cent_vs_pt}.

We see in \cref{fig:RpA_vs_centrality_pPb_502TeV} that IP-Glasma describes both the magnitude and centrality dependence of the measured $R_{pA}$ remarkably well,
despite being constrained only by HERA data.
Quantitatively,
the agreement is best over the $5\text{--}60\%$ centrality range,
with deviations from data of $\lesssim 10\%$,
and the agreement is worse in both the most central and most peripheral events.
In the $80\text{--}100\%$ centrality class,
IP-Glasma is $100\%$ larger than the data,
which is qualitatively consistent with the missing jet-veto bias.
In the $0\text{--}5\%$ centrality class,
IP-Glasma is $25\%$ larger than the data,
although IP-Glasma and data are consistent within the theoretical and experimental uncertainties.
The enhancement of the initial-state-only IP-Glasma prediction compared to data in central \ppb collisions suggests that final-state effects, most notably energy loss, could account for the remaining suppression needed to describe the data.

The centrality dependence of $\tilde{R}_{pA}$ displayed in \cref{fig:RpA_vs_centrality_pPb_502TeV} is robust against the details of the calculation:
\cref{sec:dependence_of_results_on_numerical_choices} shows that
varying the transverse box size, the lattice spacing,
the scale at which the coupling runs, and the fragmentation function set
shifts the overall normalization of $R_{pA}$ by ${\lesssim}\,10\%$ for almost all variations.
The one effect that quantitatively changes the results is allowing the coupling to run with $\tilde{\mu} = Q_s^{\text{max}}$. Doing so increases $R_{pA}$ in the most central bin by ${\sim}30\%$, while the results in the most peripheral bin are mostly unchanged.
However, as discussed in \cref{sec:nuclear_modification_factor} and shown in \cref{fig:sensitivity_grid}, the minimum-bias $R_{pA}$ then displays a strong Cronin-like peak that is not present in the data.
Therefore, 
we expect that a coupling running with $Q_s^{\text{max}}$ is disfavored by data.

\Cref{fig:rppb_corrected_cent_vs_pt} shows the $p_T$ dependence of $\tilde{R}_{pA}$ in each centrality class.
For $p_T \lesssim 3~\mathrm{GeV}$,
IP-Glasma reproduces the data only qualitatively,
as in minimum-bias collisions,
likely because fragmentation functions are not valid in this regime.
For $p_T \gtrsim 3~\mathrm{GeV}$,
the IP-Glasma result rises slowly with $p_T$,
which we attribute to a lattice artifact:
at finer lattice spacings,
$R_{pA}$ becomes independent of $p_T$ at high $p_T$
(see \cref{sec:dependence_of_results_on_lattice_spacing}).
Up to this artifact,
IP-Glasma is consistent with the approximately $p_T$-independent $R_{pA}$ measured in the $0\text{--}60\%$ centrality classes.
In the $60\text{--}80\%$ and $80\text{--}100\%$ classes, however,
the measured $R_{pA}$ decreases with increasing $p_T$ for $p_T \gtrsim 2~\mathrm{GeV}$,
whereas the IP-Glasma result decreases only up to $p_T \simeq 5~\mathrm{GeV}$
before rising slightly.
The difference in $p_T$ dependence is consistent with the missing jet-veto bias \cite{ALICE:2014xsp}.
A higher-$p_T$ hadron typically originates from a more energetic jet,
which radiates more and therefore deposits more particles at midrapidity.
The event is then more likely to migrate into a more central class,
depleting the peripheral classes of high-$p_T$ hadrons.

In both \cref{fig:RpA_vs_centrality_pPb_502TeV,fig:rppb_corrected_cent_vs_pt},
the systematic theoretical uncertainty is largest in the most central and most peripheral classes,
and small in the intermediate classes.
The origin of the uncertainty differs between the two limits.
In the most central collisions,
the number of participants is nearly saturated,
and centrality selection is driven primarily by fluctuations in the proton substructure.
Accordingly,
the spread of $\tilde{R}_{pA}$ across the posterior is controlled by the parameters that govern these fluctuations,
most strongly by the width $\sigma$ of the saturation-scale fluctuations in \cref{eqn:saturation_scale_fluctuations}.
In the most peripheral collisions,
the spread is instead dominated by the rescaling of \cref{eqn:rescaled_raa},
which assumes that $R_{pA}$ is proportional to $\sigma_{\text{inel}}^{pp,\,\text{IPG}}$.
The most peripheral class is more sensitive to $\sigma_{\text{inel}}^{pp,\,\text{IPG}}$
than this assumption allows,
because $\sigma_{\text{inel}}^{pp,\,\text{IPG}}$ affects not only the number of binary collisions
but also whether a peripheral event is counted as inelastic at all.
Consequently,
the rescaling does not reduce the uncertainty in the most peripheral classes,
and in fact enlarges it.

\begin{figure}[!t]
	\centering
	\includegraphics[width=\linewidth]{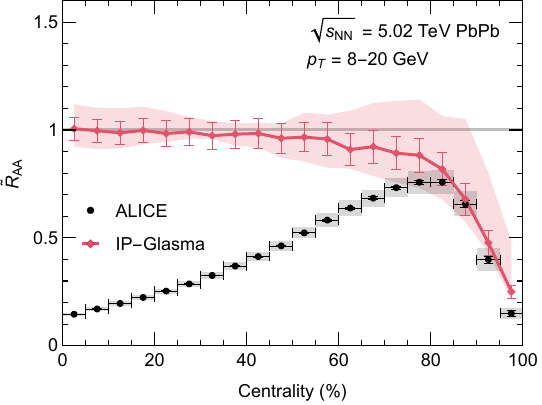}
	\caption{The rescaled nuclear modification factor $\tilde{R}_{AA}$ of \cref{eqn:rescaled_raa} for charged hadrons with $8~\mathrm{GeV} \leq p_T \leq 20~\mathrm{GeV}$ produced in \pbpb collisions at $\sqrt{s_{NN}} = 5.02~\mathrm{TeV}$ as a function of centrality. IP-Glasma results are shown in red and experimental data from ALICE~\cite{ALICE:2018ekf} are shown in black. Red bars are statistical uncertainties and bands are systematic theoretical uncertainties due to sampling from the Bayesian posterior. Black bars are statistical uncertainties and shaded gray boxes are systematic uncertainties. The unrescaled $R_{AA}$ is shown in \cref{fig:RAA_PbPb_cent_unrescaled}.}
	\label{fig:RAA_PbPb_cent}
\end{figure}

\Cref{fig:RAA_PbPb_cent} plots the rescaled nuclear modification factor $\tilde{R}_{AA}$ as a function of centrality
for $8~\mathrm{GeV} \leq p_T \leq 20~\mathrm{GeV}$ charged hadrons
produced in $\sqrt{s_{NN}} = 5.02~\mathrm{TeV}$ \pbpb collisions.
In central collisions,
the experimental data are suppressed due to jet quenching.
Final-state energy loss is not included in our model,
and therefore we find $\tilde{R}_{AA} \simeq 1 \pm 0.1$ over $0\text{--}60\%$ centrality,
consistent with binary scaling
and a factor of ${\sim}\,7$ above the data in the most central class.
In peripheral collisions, however,
the picture is more striking:
IP-Glasma captures the sharp decrease in $\tilde{R}_{AA}$ toward $80\text{--}100\%$ centrality qualitatively well.
Quantitatively, $\tilde{R}_{AA}$ produced by IP-Glasma is ${\sim} 10\%$ larger than the data in $80\text{--}85\%$ centrality collisions and ${\sim}65\%$ larger than the data in $95\text{--}100\%$ centrality collisions.
The enhancement of $\tilde{R}_{AA}$ produced by IP-Glasma compared to data in the most peripheral classes is likely due to the missing jet-veto bias;
$dN_{\text{ch}} / d\eta \simeq 22$ in $80\text{--}85\%$ centrality \pbpb collisions and $dN_{\text{ch}} / d\eta \simeq 3$ in $95\text{--}100\%$ centrality \pbpb collisions \cite{ALICE:2018ekf},
comparable to the $\mathcal{O}(5)$ particles produced by jets.
In semi-peripheral classes the disagreement may also be affected by residual energy loss \cite{Faraday:2024qtl,Faraday:2025pto}.

\begin{figure}[!htbp]
	\centering
	\includegraphics[width=\linewidth]{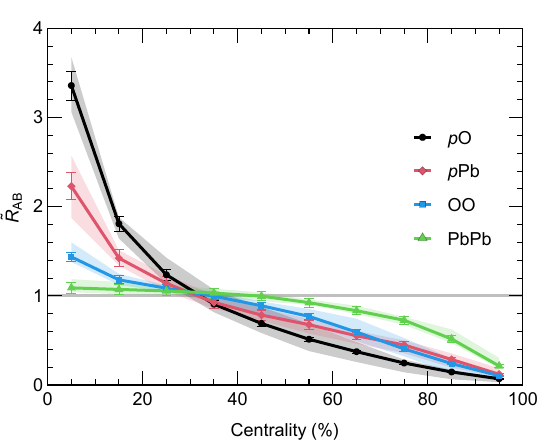}
	\caption{The rescaled nuclear modification factor $\tilde{R}_{AB}$ of \cref{eqn:rescaled_raa} for charged hadrons produced in a variety of collision systems as a function of centrality. IP-Glasma results are shown for \po collisions at $\sqrt{s_{NN}} = 9.62~\mathrm{TeV}$ (black circles), \ppb collisions at $\sqrt{s_{NN}} = 5.02~\mathrm{TeV}$ (red diamonds), \oo collisions at $\sqrt{s_{NN}} = 5.36~\mathrm{TeV}$ (blue squares), and \pbpb collisions at $\sqrt{s_{NN}} = 5.02~\mathrm{TeV}$ (green triangles). All $\tilde{R}_{AB}$ are computed with the value of $\Ncoll$ from IP-Glasma, and are rescaled sample by sample as in \cref{fig:RpA_vs_centrality_pPb_502TeV,fig:RAA_PbPb_cent}, taking for $\sigma_{\text{inel}}^{NN,\,\text{Glauber}}$ the cross section entering the Glauber model from which that $\Ncoll$ is obtained: $67~\mathrm{mb}$ for \ppb and \pbpb at $\sqrt{s_{NN}} = 5.02~\mathrm{TeV}$, $70~\mathrm{mb}$ for \oo at $\sqrt{s_{NN}} = 5.36~\mathrm{TeV}$, and $74.6~\mathrm{mb}$ for \po at $\sqrt{s_{NN}} = 9.62~\mathrm{TeV}$. Bars are statistical uncertainties, bands are systematic theoretical uncertainties due to sampling from the Bayesian posterior. The unrescaled $R_{AB}$ is shown in \cref{fig:RAA_system_comparison_unrescaled}.}
	\label{fig:RAA_system_comparison}
\end{figure}

In \cref{fig:RAA_system_comparison} we plot the rescaled $\tilde{R}_{AB}$ of \cref{eqn:rescaled_raa} from IP-Glasma for charged hadrons produced in
$\sqrt{s_{NN}} = 9.62 ~\mathrm{TeV}$ \po collisions (black),
$\sqrt{s_{NN}} = 5.02 ~\mathrm{TeV}$ \ppb collisions (red),
$\sqrt{s_{NN}} = 5.36 ~\mathrm{TeV}$ \oo collisions (blue),
and $\sqrt{s_{NN}} = 5.02 ~\mathrm{TeV}$ \pbpb collisions (green)
as a function of centrality.
Since the experimentally reported number of binary collisions is not available for \po or \oo collisions, we use the $\Ncoll$ from IP-Glasma for all four systems in this figure.

In \cref{fig:RAA_system_comparison},
we see that deviations of $\tilde{R}_{AB}$ are ordered by system size:
\pbpb, \oo, \ppb, and \po
yield $\tilde{R}_{AB}$ of ${\sim}1.05, 1.4, 2.2,$ and $3.4$
in $0\text{--}5\%$ centrality collisions, respectively.
The systems remain approximately ordered in their absolute deviation of $\tilde{R}_{AB}$ from unity over the entire centrality range,
although the differences between systems are much larger in central than in peripheral collisions.
The asymmetry between central and peripheral collisions is mostly an artifact of $\tilde{R}_{AB}$ being a ratio:
the effect size is proportionally similar in peripheral and central classes.
Another interesting feature of \cref{fig:RAA_system_comparison} is that all systems yield $\tilde{R}_{AB} \sim 1$ at ${\sim}30\%$ centrality,
although we do not know of a mechanism for this effect.

\section{Physical mechanism underlying hard-soft correlations}
\label{sec:physical_mechanism_that_leads_to_hardsoft_correlations}

In \cref{sec:centralitydependent_nuclear_modification_factor}
we showed that IP-Glasma describes the measured centrality dependence of $R_{AB}$
in small systems and in peripheral large systems.
In this section we show that the centrality dependence
follows from a single feature of the model:
soft and semi-hard particles are produced by the same classical color fields,
so that an event with an upward fluctuation in soft multiplicity
also has an upward fluctuation in hard yield.
Because centrality classes are defined by the soft multiplicity,
selecting a centrality class also changes the hard yield,
even among events with the same number of binary collisions.

All results in this section use the parameter set
$m = 0.2~\mathrm{GeV}$,
$N_q = 3$,
$B_{qc} = 4.0~\mathrm{GeV}^{-2}$,
$B_q = 0.3~\mathrm{GeV}^{-2}$,
$\sigma = 0.6$,
$C^{-1} = 0.8$,
and $d_{q,\text{min}} = 0~\mathrm{fm}$.
These values were used in prior work~\cite{Schenke:2013dpa,Schenke:2020mbo}
and lie close to the MAP values from the Bayesian posterior~\cite{Mantysaari:2022ffw}.

\begin{figure}[!t]
  \centering
  \includegraphics[width=\linewidth]{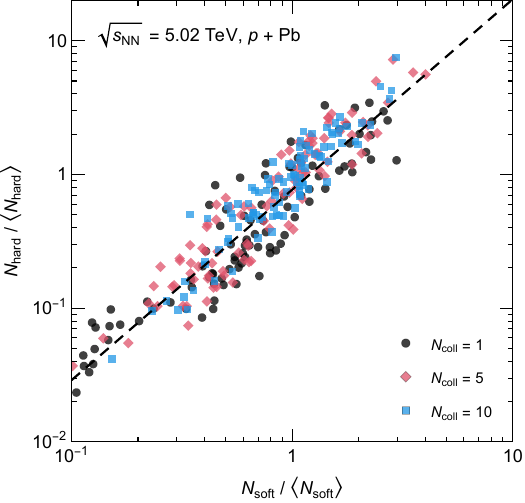}
  \caption{The self-normalized hard yield $N_{\text{hard}} / \avg{N_{\text{hard}}}$
  against the self-normalized soft yield $N_{\text{soft}} / \avg{N_{\text{soft}}}$
  in $\sqrt{s_{NN}} = 5.02~\mathrm{TeV}$ \ppb collisions.
  Each quantity is normalized by its mean within the corresponding fixed-$\Ncoll$ selection,
  so that every event sample is centered at $(1,1)$.
  Colors and markers indicate $\Ncoll = 1$ (black circles),
  $\Ncoll = 5$ (red diamonds), and $\Ncoll = 10$ (blue squares).
  The dashed line is a power-law fit to the three samples combined.
  Soft and hard hadrons are defined by $p_T < 2~\mathrm{GeV}$
  and $p_T > 4~\mathrm{GeV}$, respectively.}
  \label{fig:hard_vs_soft_fixed_ncoll}
\end{figure}

\Cref{fig:hard_vs_soft_fixed_ncoll} plots the number of hard ($p_T > 4 ~\mathrm{GeV}$) hadrons $N_{\text{hard}}$ normalized by the average hard yield $\left\langle N_{\text{hard}} \right\rangle$ against the number of soft ($p_T < 2 ~\mathrm{GeV}$) hadrons normalized by the average soft yield $\left\langle N_{\text{soft}} \right\rangle$.
Results are shown for $\sqrt{s_{NN}} = 5.02 ~\mathrm{TeV}$ \ppb collisions.
Three sets of data are shown with fixed \Ncoll and $N_{\text{part}} = N_{\text{coll}} + 1$;
each dataset is normalized by its own mean.
We observe in the figure a clear correlation between fluctuations in the number of soft particles produced and the number of hard particles produced.
Since $\Ncoll$ is fixed within each sample,
the correlation cannot arise from the geometric variation
that the Glauber model already accounts for.
We find empirically from the line of best fit in \cref{fig:hard_vs_soft_fixed_ncoll}
that $N_{\text{hard}} \propto N_{\text{soft}}^{1.3}$ at fixed $N_{\text{coll}}$ and $N_{\text{part}}$ with a correlation coefficient of $r = 0.9$.
We may already infer from \cref{fig:hard_vs_soft_fixed_ncoll}
that an upward fluctuation in the soft yield
biases an event toward a more central class
while simultaneously increasing the hard yield, resulting in $R_{AB} > 1$.
Conversely, a downward fluctuation in the soft yield biases an event toward a more peripheral centrality class, resulting in $R_{AB} < 1$.
These features qualitatively account for the centrality dependence of $R_{pA}$ observed in \cref{fig:RpA_vs_centrality_pPb_502TeV}.

The origin of the correlation between soft and hard particle production in IP-Glasma is their common dependence on the underlying color fields.
The strength of the color fields varies event by event due to various physical effects described in \cref{sec:model}:
fluctuations in the saturation scale,
the sampling of the color charges,
and the positions of the hotspots.
Rather than tracking each of these microscopic fluctuations separately, we seek a simple event-level quantity that captures their combined effect on particle production.

A natural motivation comes from the approximate scaling of the total gluon multiplicity in the CGC~\cite{Krasnitz:2000gz},
\begin{equation}
N^g \sim Q_s^2 \, \Sperp,
\label{eqn:kln}
\end{equation}
where $\Sperp$ is the transverse overlap area.
\Cref{eqn:kln} suggests that the combination $Q_s^2 \Sperp$ provides a useful measure of the overall strength of the color fields in an event.

In a collision of two nuclei $A$ and $B$, however,
the two projectiles generically have different local saturation scales,
$Q^2_{s,A}(\bt)$ and $Q^2_{s,B}(\bt)$, at each transverse position $\bt$,
so that a single $Q_s$ is not well defined.
Within the $k_\perp$-factorization picture of gluon production~\cite{Kharzeev:2004if},
the number of gluons produced per unit transverse area is controlled by the smaller of the two scales.
We therefore characterize the color fields of an event by
\begin{equation}
\Qsminsq \Sperp \equiv \int d^2 b_\perp \;
\min\!\left[ Q^2_{s,A}(\bt),\, Q^2_{s,B}(\bt) \right],
\label{eqn:qsmin_st}
\end{equation}
which in practice is evaluated as a sum over the lattice cells that carry a nonzero color field.

\begin{figure}[!t]
  \centering
  \includegraphics[width=\linewidth]{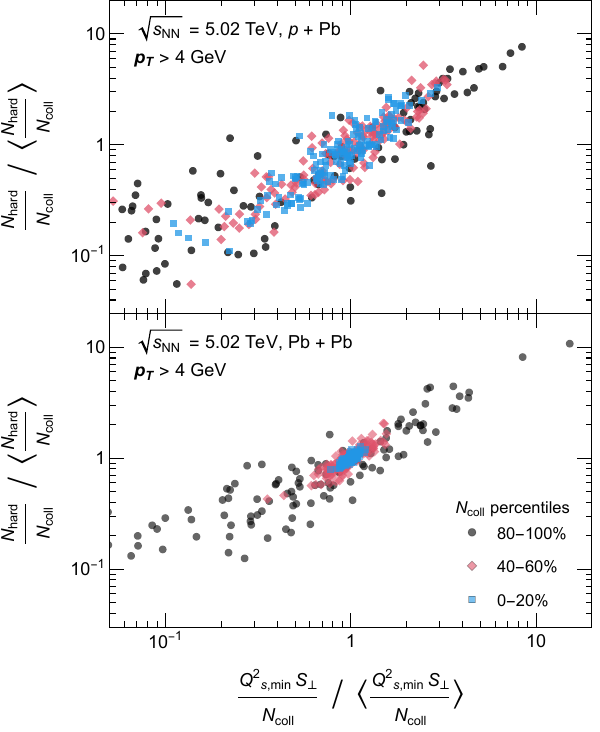}
  \caption{The self-normalized hard yield per binary collision
  $(N_{\text{hard}}/\Ncoll) / \avg{N_{\text{hard}}/\Ncoll}$
  against the self-normalized saturation scale
  $(\Qsminsq \Sperp/\Ncoll) / \avg{\Qsminsq \Sperp/\Ncoll}$
  of \cref{eqn:qsmin_st},
  where $Q_{s,\text{min}}$ is the local pointwise minimum saturation scale
  of the two colliding nuclei and $\Sperp$ is the transverse overlap area.
  The upper panel shows $\sqrt{s_{NN}} = 5.02~\mathrm{TeV}$ \ppb collisions
  and the lower panel \pbpb collisions at the same energy,
  for charged hadrons with $p_T > 4~\mathrm{GeV}$.
  Colors and markers denote percentiles of $\Ncoll$:
  $80\text{--}100\%$ (black circles),
  $40\text{--}60\%$ (red diamonds),
  and $0\text{--}20\%$ (blue squares).}
  \label{fig:self_norm_fluctuations_combined}
\end{figure}

\Cref{fig:self_norm_fluctuations_combined} plots the self-normalized hard yield
against the self-normalized $\Qsminsq \Sperp$
for \ppb collisions in the upper panel
and \pbpb collisions in the lower panel.
Both quantities are divided by $\Ncoll$
in order to separate the subnucleonic fluctuations from the overall binary scaling,
and events are binned by $\Ncoll$ percentile rather than by multiplicity-based centrality
so that the samples are unbiased%
\footnote{To avoid artifacts from the discrete nature of $\Ncoll$,
we apply a small stochastic perturbation to $\Ncoll$ in each event.}.
The hard yield rises with $\Qsminsq \Sperp$ in both systems,
demonstrating that fluctuations in the strength of the color fields
account for a substantial portion of the correlation in \cref{fig:hard_vs_soft_fixed_ncoll}.
The remaining variance not captured by $\Qsminsq S_{\perp}$ in \cref{fig:self_norm_fluctuations_combined} is then likely attributable to the other subnucleonic fluctuations in the model discussed at the start of the section.

Two quantities therefore control the magnitude of the bias in $R_{AB}$:
the slope with which the hard yield responds to a fluctuation in $\Qsminsq \Sperp$,
and the width of the residual fluctuations in $\Qsminsq \Sperp$ at fixed $\Ncoll$.
The slope is approximately independent of $\Ncoll$
and is the same in the two collision systems,
so it is the width of the $\Qsminsq \Sperp / \Ncoll$ distribution, visible in \cref{fig:self_norm_fluctuations_combined} as the horizontal extent of each sample,
that distinguishes one centrality class and one collision system from another.

In \ppb collisions the relative width falls by only a factor of ${\sim}3$,
from ${\sim}1.6$ in the $80\text{--}100\%$ \Ncoll percentile class to ${\sim}0.5$ in the $0\text{--}20\%$ \Ncoll percentile class,
and so remains of order unity even in the most central collisions.
The centrality bias is therefore present in every \ppb centrality class,
which is why $R_{p\pb}$ in \cref{fig:RpA_vs_centrality_pPb_502TeV}
depends so strongly on centrality.
In \pbpb collisions,
the relative width falls much further,
from ${\sim}1.5$ in the $80\text{--}100\%$ $\Ncoll$ percentile class to ${\sim}0.06$ in the $0\text{--}20\%$ class. 
The resulting small fluctuations in central collisions and large fluctuations in peripheral collisions explain the $R_{AA} \sim 1$ in central classes and $R_{AA} \ll 1$ in peripheral classes.
The reduction in the width in more central events reflects the increasing number of independent subnucleonic sources contributing to the collision. 
In central \pbpb collisions, fluctuations associated with individual nucleons and their substructure are averaged over many participants, reducing their relative contribution to the event-by-event variation in $\Qsminsq \Sperp / \Ncoll$.

\FloatBarrier
\section{Conclusions}
\label{sec:conclusions}

We presented predictions from the IP-Glasma model in which all particles%
---both soft and semi-hard---%
are produced within a single theoretical framework.
We found that,
constrained by a Bayesian analysis using only HERA data,
the IP-Glasma model describes the hard-soft correlations in high-energy hadronic collisions qualitatively and, in several observables, quantitatively.

We first validated the model against self-normalized multiplicity distributions,
showing that IP-Glasma agrees with data from \pp, \po, \ppb, \oo, and \pbpb
collisions over the ${\sim}1\text{--}100\%$ centrality range to within ${\sim}30\%$,
and within the combined theoretical and experimental systematic uncertainties.
The exception is the rare high-multiplicity tail beyond the ultracentral $1\%$ centrality class, which is challenging for most models to describe~\cite{ALICE:2022xip,ATLAS:2026xcm},
and is overpredicted by IP-Glasma. 

IP-Glasma does not simultaneously describe
the absolute value of the transverse momentum spectrum for both
soft ($p_T \lesssim 2 ~\mathrm{GeV}$) and semi-hard ($p_T \sim 4\text{--}20 ~\mathrm{GeV}$) particles. However, the normalization can be chosen such that the high-$p_T$ part of the spectrum is well described, allowing for studies of $R_{AB}$.
Our model predicts $R_{pA} \sim 1.2 \pm 0.3$ for minimum-bias $\sqrt{s_{NN}} = 5.02 ~\mathrm{TeV}$ \ppb collisions for $p_T \gtrsim 5 ~\mathrm{GeV}$,
in good agreement with experimental data \cite{ATLAS:2022kqu,ALICE:2014xsp}, albeit with large uncertainties.
For $p_T \lesssim 5 ~\mathrm{GeV}$, 
our model predicts less suppression than observed in the experimental data, likely due to the inapplicability of fragmentation functions at such small momenta.

The large theoretical uncertainty, originating from sampling parameters from the posterior distribution of the Bayesian analysis, can be reduced significantly by allowing only those parameter sets that reproduce the correct inelastic nucleon-nucleon cross section.

We presented predictions for the minimum-bias nuclear modification factor
$R^{\sigma}_{AB}$ of \cref{eqn:nuclear_modification_factor_sigma},
in which the normalization uncertainty associated with the inelastic cross section largely cancels,
as a function of $p_T$ in \po and \oo collisions.
We found good agreement with the ALICE data in \po collisions,
while the IP-Glasma calculation could not reproduce the suppression observed in \oo collisions, 
as expected if the observed $R^{\sigma}_{OO}<1$ arises from final-state energy loss~\cite{Faraday:2025pto}.

We presented predictions for the centrality-cut $R_{p\pb}$,
correcting each posterior sample for the mismatch between its \pp inelastic cross section and the one used in the Glauber model,
and compared to experimental data~\cite{ALICE:2014xsp}.
We found that IP-Glasma reproduces the trend observed in data remarkably well:
$R_{p\pb}$ is strongly enhanced in central collisions and strongly suppressed in peripheral collisions. We found quantitative disagreement in the most peripheral collisions, which could arise from a ``jet-veto bias''~\cite{ALICE:2014xsp} that is absent from our model; the growth of the disagreement with $p_T$ is consistent with this interpretation. Further, the ${\sim}25\%$ enhancement of the IP-Glasma prediction over the data in the most central class leaves room for additional suppression from final-state energy loss.

We further studied the centrality-dependent $R_{AA}$ in $\sqrt{s_{NN}} = 5.02 ~\mathrm{TeV}$ \pbpb collisions. As expected, the model does not reproduce the measured suppression in $0\text{--}60\%$ central collisions, as we do not include jet quenching.
For $80\text{--}100\%$ centrality \pbpb collisions our model shows a sharp decrease in $R_{AA}$ as a function of centrality, in qualitative agreement with experimental data \cite{ALICE:2018ekf}. 

Comparing the centrality-dependent $R_{AB}$ across all four collision systems,
we found a smooth ordering with system size:
the deviation from binary scaling grows as the collision system shrinks,
running monotonically from \pbpb through \oo and \ppb to \po.

We also presented a physical mechanism for the hard-soft correlations in IP-Glasma that accounts for their ordering with system size and centrality: even at fixed $N_{\rm coll}$, fluctuations in $Q_s$ and $\Sperp$ drive correlated fluctuations in hard- and soft-particle production. In small and peripheral large systems, where few nucleons participate, these fluctuations can significantly shift an event's centrality class relative to Glauber-model expectations.

This mechanism parallels how deviations of $R_{AB}$ from unity have been described using PYTHIA coupled with the Glauber model \cite{ALICE:2014xsp,Loizides:2017sqq}, where the correlation arises because both hard and soft production depend on the number of multi-parton interactions (MPIs). Our approach offers four advantages over PYTHIA + Glauber.
First, IP-Glasma combines a first-principles description of the dynamics with a phenomenological treatment of geometry in a single self-consistent framework.
Second, the model parameters are constrained by HERA data without additional tuning,
so that our predictions are parameter-free apart from the overall normalization and the scale at which the coupling runs.
Third, IP-Glasma captures both the approximate participant scaling at low $p_T$~\cite{Schenke:2013dpa} and the approximate binary scaling at high $p_T$ within a single calculation,
whereas a naive PYTHIA + Glauber construction that superimposes $N_{\rm coll}$ independent PYTHIA events gives binary scaling at all $p_T$~\cite{ALICE:2014xsp,Loizides:2017sqq}.
Finally, because the hard partons arise from the high-$k_T$ tail of the same classical color fields that generate the soft sector,
IP-Glasma is a natural starting point for a framework unifying jet quenching, hydrodynamic evolution, and hard-soft correlations.

Several natural extensions of this framework remain for future work.
The most immediate is to constrain the definition of an inelastic collision and the hotspot size parameter $B_q$
using the measured inelastic nucleon-nucleon cross section.
As shown in this work,
such a constraint would substantially reduce the theoretical uncertainty on the normalization of $R_{AB}$.

The hardness of the high-$p_T$ spectrum compared to data
is a generic feature of the McLerran-Venugopalan (MV) model \cite{McLerran:1993ka, McLerran:1994vd}
and can be remedied by introducing an anomalous dimension that softens the spectrum \cite{Dumitru:2005gt, ALbacete:2010ad, Tribedy:2010ab}.
At low $p_T$,
a more realistic hadronization model~\cite{Greif:2020rhi}
would likely improve the description of $R_{AB}$.
Coupling IP-Glasma to a vacuum parton shower would enable us to capture the missing jet-veto bias, which is important for describing peripheral collisions.

A more comprehensive extension would couple IP-Glasma to hydrodynamic evolution~\cite{Schenke:2020mbo,Schenke:2010rr,Schenke:2010nt,Schenke:2011bn}
and include jet quenching~\cite{Faraday:2025pto}.
Such a program would provide a framework in which the initial-state hard-soft correlations studied here along with final-state energy loss and hydrodynamic evolution are all treated simultaneously,
allowing the low- and high-$p_T$ data in small systems to be described within a single calculation.
Although such a project would be computationally demanding,
a Bayesian analysis incorporating both HERA data and low- and high-$p_T$ hadronic-collision data from RHIC and the LHC
could constrain the parameters governing the fluctuating proton,
nuclear geometry,
hydrodynamic evolution,
and jet quenching simultaneously.

An observable of significant current interest that we have not yet discussed is the high-$p_T$ $v_2$.
Previous work has shown that it is difficult to produce a nonzero, measurable high-$p_T$ $v_2$ using energy loss models \cite{Bert:2026uxa},
and so initial-state effects are a natural potential explanation for the measured large $v_2$ in \ppb collisions \cite{ATLAS:2019vcm,CMS:2025kzg}.
While we did not compute the $v_2$ in this work, preliminary investigations showed that the $v_2$ produced by IP-Glasma at high $p_T$ was an order of magnitude smaller than that required to explain the $v_2\{4\} \sim 0.1$ in central \ppb collisions, consistent with previous work \cite{Greif:2020rhi}.

Future experimental lines of investigation
include a measurement of the centrality-cut nuclear modification factor in \oo collisions.
While the minimum-bias $R_{AA}$ is a natural first measurement to perform,
as it avoids many of the selection biases and hard-soft correlations discussed in this work \cite{Huss:2020dwe,Huss:2020whe},
it would be extremely interesting to study a system where both energy loss and hard-soft correlations are present.
Such a study would be a stepping stone toward understanding jet quenching in exceptionally small systems,
where centrality selection is unavoidable if the expected energy loss signal is to be large enough to measure.
More broadly,
experiments should report high-$p_T$ results using multiple centrality definitions.
Centrality defined by percentiles of midrapidity multiplicity is especially useful: although sensitive to hard-soft correlations and selection biases, a midrapidity estimator is calculable in boost-invariant frameworks and less sensitive to long-range correlation structures, which remain poorly understood.
Complementary paths toward studying energy loss in extremely small systems include collisions of small but symmetric systems such as helium or lithium,
where sensitivity to fluctuations in the incoming proton should be reduced compared to \pa collisions~\cite{Faraday:2025prr}.

\section*{Acknowledgments}

The authors are grateful for discussions with Matthew D.\ Sievert, Vladimir Skokov, and Nicolas Strangmann. Computations were performed using facilities provided by the University of Cape Town's ICTS High Performance Computing team: \href{http://hpc.uct.ac.za}{hpc.uct.ac.za}. The authors acknowledge the Centre for High Performance Computing (CHPC), South Africa, for providing computational resources to this research project.
CF and WAH thank the National Research Foundation and the SA-CERN collaboration for their generous financial support during this work. CF thanks the University of Cape Town for financial support during this work. This research was conducted in part by WAH while visiting the Okinawa Institute of Science and Technology (OIST) through the Theoretical Sciences Visiting Program (TSVP). 
This work is supported by the U.S. Department of Energy, Office of Science, Office of Nuclear Physics, under DOE Contract No.~DE-SC0012704 and within the framework of the Saturated Glue (SURGE) Topical Theory Collaboration (BPS).

\appendix
\crefalias{section}{appendix}
\crefalias{subsection}{appendix}

\section{Dependence of results on numerical and physical choices}
\label{sec:dependence_of_results_on_numerical_choices}

\begin{figure*}[!p]
  \centering
  \begin{subfigure}{\linewidth}
    \includegraphics[width=\linewidth]{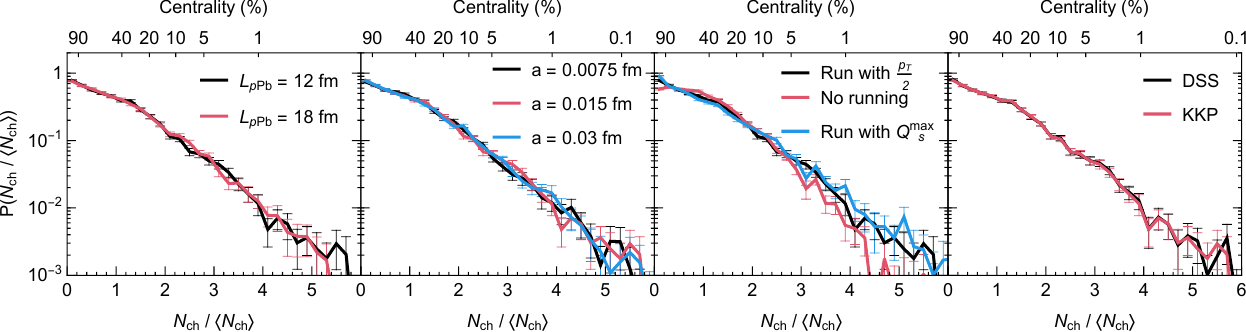}
    \caption{The self-normalized multiplicity distribution $P(\Nch / \avg{\Nch})$ as a function of $\Nch / \left\langle \Nch \right\rangle$.}
    \label{fig:sensitivity_pnch}
  \end{subfigure}
  \\[0.5ex]
  \begin{subfigure}{\linewidth}
    \includegraphics[width=\linewidth]{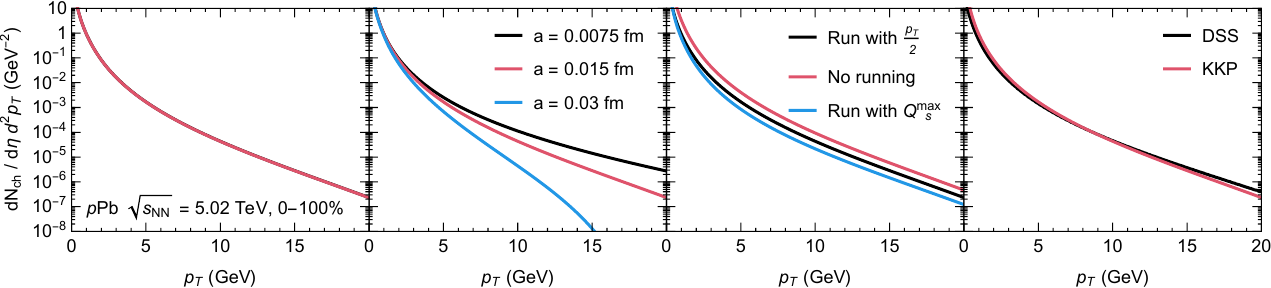}
    \caption{The charged hadron spectrum $d^3 \Nch / d^2 p_T\, d\eta$ as a function of $p_T$.}
    \label{fig:sensitivity_spectrum}
  \end{subfigure}
  \\[0.5ex]
  \begin{subfigure}{\linewidth}
    \includegraphics[width=\linewidth]{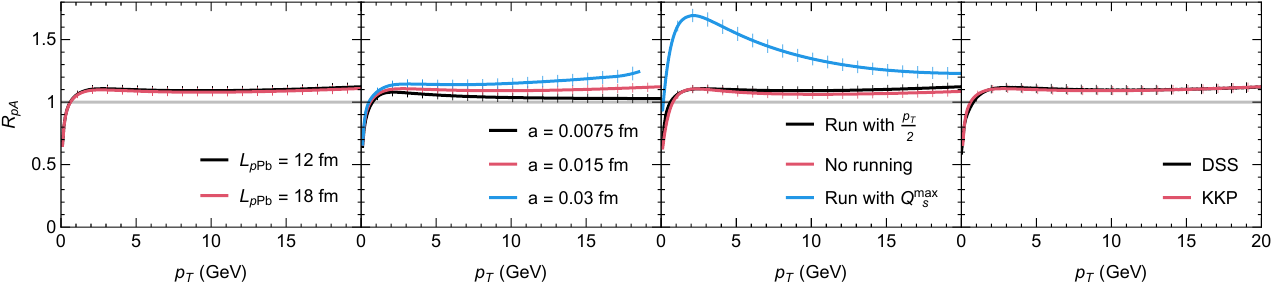}
    \caption{The minimum-bias nuclear modification factor $R_{pA}$ as a function of $p_T$.}
    \label{fig:sensitivity_rpamb}
  \end{subfigure}
  \\[0.5ex]
  \begin{subfigure}{\linewidth}
    \includegraphics[width=\linewidth]{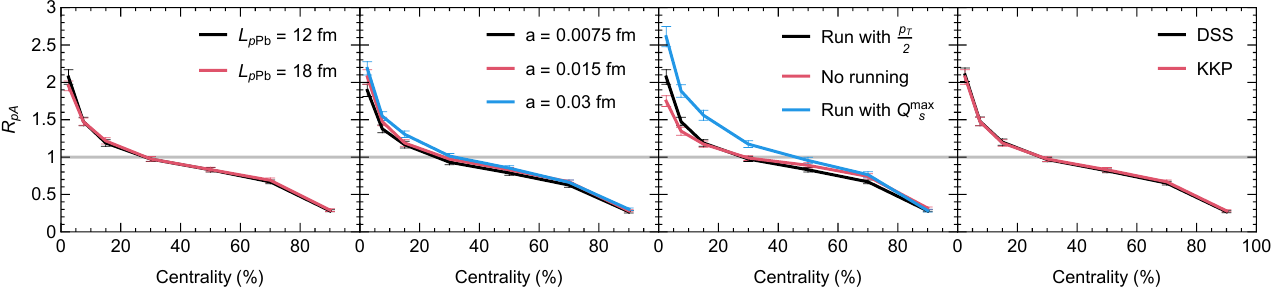}
    \caption{The nuclear modification factor $R_{pA}$ as a function of centrality.}
    \label{fig:sensitivity_rpacent}
  \end{subfigure}
  \caption{Sensitivity of the four main observables used in this work to the numerical and physical choices entering the calculation, for \ppb collisions at $\sqrt{s_{NN}} = 5.02 ~\mathrm{TeV}$. Within each of panels \subref{fig:sensitivity_pnch}--\subref{fig:sensitivity_rpacent} the columns show, from left to right, the variation of the transverse box size $L$, the lattice spacing $a$, the scale $\tilde{\mu}$ at which the coupling runs, and the fragmentation function set, with all parameters other than the one being varied held at their default values. Bars are statistical uncertainties.}
  \label{fig:sensitivity_grid}
\end{figure*}

We will now investigate the dependence of our results presented in \cref{sec:validation,sec:nuclear_modification_factor,sec:centralitydependent_nuclear_modification_factor} on various details of the simulation. The main numerical parameters that the simulation depends on are the size $L$ of the transverse box and the lattice spacing $a$. Additionally, there are physical choices that we have made, most notably the scale at which the coupling runs and the hadronization procedure. In this section we do not sample from the Bayesian posterior, as was done in the main body of the text. Instead, we compute all results for a canonical choice of parameters: %
$m = 0.2 ~\mathrm{GeV}$,
$N_q = 3$,
$B_{qc} = 4.0 ~\mathrm{GeV}^{-2}$,
$B_q = 0.3 ~\mathrm{GeV}^{-2}$,
$\sigma = 0.6$,
$C^{-1} = 0.8$,
and $d_{q,\text{min}} = 0 ~\mathrm{fm}$, %
which lie close to the MAP values quoted in \cref{sec:bayesian_posterior_sampling}.
The uncertainties shown in this appendix are therefore statistical only.
Because a single parameter set is used rather than the posterior,
the absolute values quoted in this appendix
are not directly comparable to the posterior-averaged results in the body;
only the differences between the variations considered here are meaningful.
All results in this section are for \ppb collisions at $\sqrt{s_{NN}} = 5.02 ~\mathrm{TeV}$,
and each variation is performed with all other parameters held fixed at their default values of
$L = 12 ~\mathrm{fm}$,
$a = 0.015 ~\mathrm{fm}$,
$\tilde{\mu} = p_T/2$,
and the DSS fragmentation functions \cite{deFlorian:2007aj}.
\Cref{fig:sensitivity_grid} displays the effect of all the aforementioned variations on the four main observables considered in this work. 

\subsection{Box size and lattice spacing}
\label{sec:dependence_of_results_on_lattice_spacing}

The first column of \cref{fig:sensitivity_grid} shows that
varying the transverse box size $L$ from $12 ~\mathrm{fm}$ to $18 ~\mathrm{fm}$
changes every observable by less than $1\%$.
We performed the same check for \pp collisions,
which use the smallest box in this work, $L = 6 ~\mathrm{fm}$,
and which provide the denominator of every $R_{AB}$ reported here:
repeating the calculation at $L = 9 ~\mathrm{fm}$ shows a similarly small effect.

The lattice spacing $a$ (second column of each panel in \cref{fig:sensitivity_grid}) is a far more important choice,
since the spacing sets the ultraviolet cutoff of the simulation.
Modes with transverse momentum approaching the edge of the Brillouin zone
are increasingly distorted by the lattice dispersion relation,
so we expect reliable results only for
$p_T \lesssim (\pi/2)/a$,
which corresponds to
$p_T \lesssim 10$, $21$, and $41 ~\mathrm{GeV}$
for $a = 0.03$, $0.015$, and $0.0075 ~\mathrm{fm}$, respectively.
Since the observables of interest in this work lie in the range
$5 ~\mathrm{GeV} \lesssim p_T \lesssim 20 ~\mathrm{GeV}$,
this cutoff is not comfortably far away,
and indeed the spectrum is the observable most strongly affected:
the three curves agree for $p_T \lesssim 3 ~\mathrm{GeV}$,
but the coarser lattices fall progressively more steeply above this scale,
with the $a = 0.03 ~\mathrm{fm}$ result falling below the
$a = 0.0075 ~\mathrm{fm}$ result by two orders of magnitude
already at $p_T \simeq 12 ~\mathrm{GeV}$.

The onset of the deviation in each case tracks the corresponding
lattice cutoff $(\pi/2)/a$, as expected.
The severity of the effect is amplified by the hadronization step.
Because \cref{eqn:hadron_spectrum} convolves the gluon spectrum with the fragmentation function over $z$,
a hadron of momentum $p_T^h$ receives contributions from gluons of momentum $p_T^g = p_T^h / z > p_T^h$.
The lattice supports no gluons above the Brillouin-zone edge,
so this convolution is truncated unphysically at small $z$, for $z \lesssim (2/\pi)\, p_T^h a$,
and the hadron spectrum is depleted at momenta well below the nominal gluon-level cutoff $(\pi/2)/a$.
Combined with the steepness of the spectrum,
this truncation explains why the spectrum produced by coarser lattices falls short of that produced by finer lattices by orders of magnitude.

Much of the dependence on $a$ cancels in $R_{pA}$,
which differs between the three spacings by
${\lesssim}\,25\%$.
The cancellation is not complete, however.
For $p_T \gtrsim 10 ~\mathrm{GeV}$,
the coarser lattices produce an $R_{pA}$ that rises with $p_T$,
reaching $R_{pA} \sim 1.25$ at $p_T = 20 ~\mathrm{GeV}$
for $a = 0.03 ~\mathrm{fm}$,
while the finest lattice, $a = 0.0075 ~\mathrm{fm}$,
gives a flat $R_{pA} \simeq 1$ out to $p_T = 20 ~\mathrm{GeV}$.
The behavior of the $a = 0.0075 ~\mathrm{fm}$ result is a meaningful check
on our framework:
analytically, one expects $R_{pA} \to 1$ at large $p_T$ \cite{Kharzeev:2002pc,Kharzeev:2003wz,Lappi:2013zma},
and only the finest lattice reproduces this limit.
This lattice sensitivity also provides a potential explanation for the residual
${\sim}\,10\%$ enhancement above unity
that survives the normalization correction of \cref{sec:nuclear_modification_factor}:
that residual is consistent with a lattice artifact
at our default spacing of $a = 0.015 ~\mathrm{fm}$
rather than a physical effect.
However, confirming that the enhancement of $R_{pA}$ by ${\sim}10\%$ above one 
is due to lattice effects
would require repeating the analysis 
for different values of $a$
for all of the Bayesian posteriors considered in this work.

In the centrality dependence of $R_{pA}$, \cref{fig:sensitivity_rpacent},
the two finer spacings are nearly indistinguishable,
agreeing to within a few percent at every centrality,
while the coarsest lattice sits up to ${\sim}\,10\%$ above them
over the $5\text{--}40\%$ range
before converging in the most peripheral bins.
Since our production runs use $a = 0.015 ~\mathrm{fm}$,
we do not anticipate significant sensitivity to the lattice spacing for centrality-cut $R_{AB}$.
While we do not show the $p_T$ dependence of $R_{pA}$ within each centrality class for different lattice spacings, 
we found similar lattice artifacts as the minimum bias result.
The $R_{pA}$ rises slowly with $p_T$ for $p_T \gtrsim 6 ~\mathrm{GeV}$ for $a = 0.03$ and $a = 0.015$,
while for $a = 0.0075~\mathrm{fm}$
$R_{pA}$ is flat or slightly decreasing out to $p_T = 20~\mathrm{GeV}$.
The slight rise at high $p_T$ of the IP-Glasma result in the most peripheral classes of \cref{fig:rppb_corrected_cent_vs_pt}
is therefore a lattice artifact rather than a physical effect.

We also computed $\sigma_{\text{inel}}^{pp}$ for each variation of the box size and lattice spacing, and found that all values of $\sigma_{\text{inel}}^{pp}$ agreed within statistical uncertainties.

\subsection{Running coupling scale and fragmentation functions}
\label{sec:dependence_of_results_on_physical_choices}

The running coupling enters the gluon multiplicity through the factor
$g^2 / (4\pi \alpha_s(\tilde{\mu}))$ discussed in \cref{sec:ipglasma},
and therefore requires a choice of the scale $\tilde{\mu}$
at which the coupling is evaluated.
Throughout this work we take $\tilde{\mu} = p_T / 2$,
so that the coupling runs with the momentum of the produced gluon,
following \cite{Schenke:2013dpa}.
The third column of each panel of \cref{fig:sensitivity_grid} compares this default
against two commonly used alternatives.
The first is to switch the running off, so that $\alpha_s$ carries no scale dependence.
The second is to run instead with the larger of the two saturation scales in the collision,
$\tilde{\mu} = Q_s^{\text{max}}$.

With the running switched off,
the spectrum lies above the default by a factor of ${\sim}\,2$ (see the third column of \cref{fig:sensitivity_spectrum}),
close to an overall normalization shift
that largely cancels in the minimum-bias $R_{pA}$,
where the two agree to ${\lesssim}\,5\%$ (see the third column of \cref{fig:sensitivity_rpamb}).

Running with $Q_s^{\text{max}}$ behaves quite differently,
and is the one variation considered anywhere in this appendix
that changes the shape of an observable rather than its normalization.
As seen in the third column of \cref{fig:sensitivity_rpamb}, running with $Q_{s}^{\text{max}}$ produces a pronounced Cronin-like peak in the minimum-bias $R_{pA}$,
rising to $R_{pA} \simeq 1.7$ at $p_T \simeq 2\text{--}3 ~\mathrm{GeV}$
and falling only to $R_{pA} \simeq 1.2$ by $p_T = 20 ~\mathrm{GeV}$,
while both other prescriptions are flat at $R_{pA} \simeq 1.05\text{--}1.1$
across the entire range.
The peak is a large enough effect to be confronted directly with data.
The measured minimum-bias $R_{pA}$ in \cref{fig:RAA_pPb_MB}
rises to at most $R_{pA} \simeq 1.1\text{--}1.2$ over the entire measured range,
so a peak of $1.7$ overshoots the measurement by a wide margin,
well outside both the experimental uncertainties
and the statistical uncertainties of the calculation.
We therefore regard $\tilde{\mu} = Q_s^{\text{max}}$
as disfavored by the minimum-bias $R_{pA}$ data.

The centrality dependence of $R_{pA}$ tells the same story.
As seen in the third column of \cref{fig:sensitivity_rpacent}, running with $Q_s^{\text{max}}$ raises $R_{pA}$ in the most central bin
from $2.05$ to ${\simeq}\,2.65$,
and the curve remains above the default out to ${\sim}\,70\%$ centrality
before the three prescriptions converge in the most peripheral bins.
Switching the running off instead compresses the dynamic range,
lowering $R_{pA}$ in the most central bin
from $2.05$ to $1.75$
and raising $R_{pA}$ in the most peripheral bin
from $0.28$ to $0.32$,
while agreeing with the default to within a few percent
across the intermediate $10\text{--}60\%$ range
and crossing $R_{pA} = 1$ at the same centrality of ${\sim}\,30\%$.

The final choice is the gluon-to-hadron fragmentation function
$D_g^h(z, Q)$ entering \cref{eqn:hadron_spectrum},
for which we use the DSS set \cite{deFlorian:2007aj}.
The fourth column of each panel of \cref{fig:sensitivity_grid}
compares DSS with the Kniehl, Kramer, and P\"otter (KKP) set \cite{Kniehl:2000fe}.
We find that the differences in all observables
between results using the KKP set and the DSS set
are $\lesssim 5 \%$.

Neither the fragmentation function nor the scale at which the coupling runs can affect the inelastic cross section, and so we found that all $\sigma_{\text{inel}}^{pp}$ computed agreed within statistical uncertainties.

\section{Normalization of $R_{AB}$ and the inelastic cross section}
\label{sec:inelastic_cross_section_parameters}

\Cref{eqn:nuclear_modification_factor,eqn:nuclear_modification_factor_sigma}
define two forms of the nuclear modification factor,
and \cref{sec:observable_implementation} asserted
that the two agree only insofar as the cross sections that normalize them agree.
Here we make that statement precise
and derive \cref{eqn:rab_normalization}.

\Cref{eqn:nuclear_modification_factor,eqn:nuclear_modification_factor_sigma} differ in how they are normalized and where the normalization comes from.
$R^{\sigma}_{AB}$ is built entirely from IP-Glasma output:
both the spectra and the cross sections of \cref{eqn:inelastic_cross_section}
that convert those spectra to yields are predictions of the model.
$R_{AB}$ instead divides by an $\avg{\Ncoll}$
that IP-Glasma does not itself compute
and which we take from the Glauber analysis reported by experiment.
That Glauber analysis carries its own pair of inelastic cross sections:
a nucleon-nucleon cross section $\sigma_{\text{inel}}^{NN}$,
supplied as an input,
and a nucleus-nucleus cross section $\sigma_{\text{inel}}^{AB}$,
returned as output.
We therefore attach a superscript to every inelastic cross section below,
``IPG'' for those IP-Glasma predicts
and ``Glauber'' for those belonging to the Glauber analysis.
Nothing requires the two sets of cross sections to agree,
and we will see that the mismatch between them accounts for the entire normalization uncertainty in $R_{AB}$.

In the Glauber model the mean number of binary collisions in minimum-bias events is~\cite{Miller:2007ri}
\begin{align}
  \avg{\Ncoll} &= \frac{\int d^2 \bt \; \sigma_{\text{inel}}^{NN,\,\text{Glauber}} \, T_{AB}(\bt)}{\int d^2\bt \; P_{\text{inel}}^{AB}(\bt)} \nonumber \\
  &= AB \;\frac{\sigma_{\text{inel}}^{NN,\,\text{Glauber}}}{\sigma_{\text{inel}}^{AB,\,\text{Glauber}}} \, ,
  \label{eqn:ncoll_identity}
\end{align}
where we have used $\int d^2\bt \; T_{AB}(\bt) = AB$
together with the relation between $P_{\text{inel}}^{AB}$ and $\sigma_{\text{inel}}^{AB}$
of \cref{eqn:inelastic_cross_section}.
The definition of $R^{\sigma}_{AB}$ in \cref{eqn:nuclear_modification_factor_sigma} may be used to write
\begin{equation}
  \frac{d^3 \sigma_{\text{ch}}^{AB}}{d^2 p_T\, dy} = AB \, R^{\sigma}_{AB} \, \frac{d^3 \sigma_{\text{ch}}^{pp}}{d^2 p_T\, dy} \, ,
\end{equation}
and converting these cross sections to yields with \cref{eqn:differential_cross_section} gives
\begin{equation}
  \frac{d^3 \Nch^{AB} / d^2 p_T\, dy}{d^3 \Nch^{pp} / d^2 p_T\, dy}
  = AB \, R^{\sigma}_{AB} \, \frac{\sigma_{\text{inel}}^{pp,\,\text{IPG}}}{\sigma_{\text{inel}}^{AB,\,\text{IPG}}} \, ,
  \label{eqn:nch_in_terms_of_sigma}
\end{equation}
in which every quantity is IP-Glasma output.
Substituting \cref{eqn:ncoll_identity,eqn:nch_in_terms_of_sigma} into \cref{eqn:nuclear_modification_factor},
\begin{align}
  R_{AB} &= \frac{1}{\avg{\Ncoll}} \frac{d^3 \Nch^{AB} / d^2 p_T\, dy}{d^3 \Nch^{pp} / d^2 p_T\, dy} \nonumber \\
  &= \frac{\sigma_{\text{inel}}^{AB,\,\text{Glauber}}}{AB \, \sigma_{\text{inel}}^{NN,\,\text{Glauber}}}
  \times AB \, R^{\sigma}_{AB} \, \frac{\sigma_{\text{inel}}^{pp,\,\text{IPG}}}{\sigma_{\text{inel}}^{AB,\,\text{IPG}}} \nonumber \\
  &= R^{\sigma}_{AB}
  \times \frac{\sigma_{\text{inel}}^{pp,\,\text{IPG}}}{\sigma_{\text{inel}}^{NN,\,\text{Glauber}}}
  \times \frac{\sigma_{\text{inel}}^{AB,\,\text{Glauber}}}{\sigma_{\text{inel}}^{AB,\,\text{IPG}}} \, .
  \label{eqn:rpa_rsigma_relation}
\end{align}
The two forms of the nuclear modification factor therefore differ
by two ratios of cross sections,
each of which pairs an IP-Glasma prediction against its Glauber counterpart.

The first ratio is the dominant effect in \cref{eqn:rpa_rsigma_relation}.
The ratio compares the \pp cross section that IP-Glasma predicts
with the nucleon-nucleon cross section assumed by the Glauber analysis,
and is a single number per posterior sample,
independent of $p_T$.
The ratio is also far from unity:
$\sigma_{\text{inel}}^{pp,\,\text{IPG}}$ spans $45\text{--}145 ~\mathrm{mb}$ across the posterior
(see \cref{fig:cross_section_vs_bgq}),
a factor of three,
while $\sigma_{\text{inel}}^{NN,\,\text{Glauber}}$ is the fixed $70 ~\mathrm{mb}$ of Ref.~\cite{ALICE:2014xsp}.
This ratio is what separates \cref{fig:RAA_pPb_MB} from \cref{fig:RAA_pPb_MB_sigma}.

The second ratio differs from unity by only $\lesssim 5\%$.
For any extended system---\ppb, \oo, or \pbpb---%
$\sigma_{\text{inel}}^{AB}$ is fixed mostly by the transverse area over which nucleons from the two nuclei can overlap,
which the nuclear radii and the sampled nucleon configurations determine.
The substructure of an individual nucleon enters only near the periphery,
where a single nucleon-nucleon pair decides whether the event is inelastic,
and that periphery is a small fraction of the total area. Discarding the second ratio in \cref{eqn:rpa_rsigma_relation} yields \cref{eqn:rab_normalization}
and motivates the definition of \cref{eqn:rescaled_raa}.

\section{Unrescaled centrality-dependent \texorpdfstring{$R_{AB}$}{RAB}}
\label{sec:unrescaled_centrality}

\begin{figure}[!t]
	\centering
	\includegraphics[width=\linewidth]{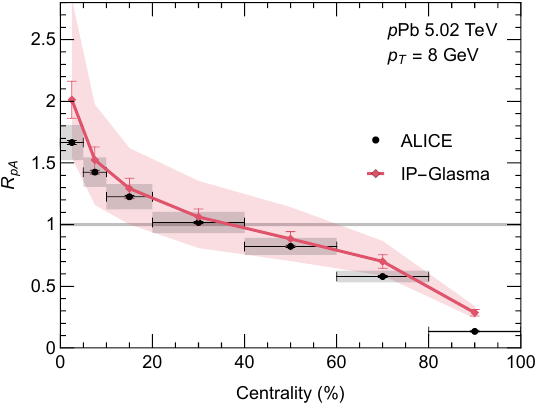}
	\caption{The unrescaled nuclear modification factor $R_{pA}$ of \cref{eqn:nuclear_modification_factor} for charged hadrons with $7.5~\mathrm{GeV} \leq p_T \leq 8.5~\mathrm{GeV}$ produced in \ppb collisions at $\sqrt{s_{NN}} = 5.02~\mathrm{TeV}$ as a function of centrality. IP-Glasma results are shown in red and experimental data from ALICE~\cite{ALICE:2014xsp} are shown in black. Red bars are statistical uncertainties, bands are systematic theoretical uncertainties due to sampling from the Bayesian posterior. Black bars are statistical uncertainties and shaded gray boxes are systematic uncertainties.}
	\label{fig:RpA_vs_centrality_pPb_502TeV_unrescaled}
\end{figure}

\begin{figure}[!t]
	\centering
	\includegraphics[width=\linewidth]{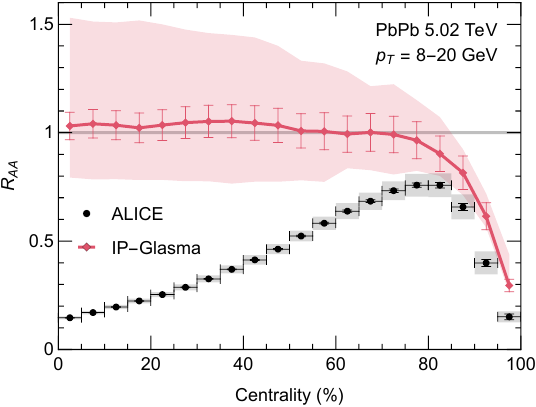}
	\caption{The unrescaled nuclear modification factor $R_{AA}$ of \cref{eqn:nuclear_modification_factor} for charged hadrons with $8~\mathrm{GeV} \leq p_T \leq 20~\mathrm{GeV}$ produced in \pbpb collisions at $\sqrt{s_{NN}} = 5.02~\mathrm{TeV}$ as a function of centrality. IP-Glasma results are shown in red and experimental data from ALICE~\cite{ALICE:2018ekf} are shown in black. Red bars are statistical uncertainties, bands are systematic theoretical uncertainties due to sampling from the Bayesian posterior. Black bars are statistical uncertainties and shaded gray boxes are systematic uncertainties.}
	\label{fig:RAA_PbPb_cent_unrescaled}
\end{figure}

\begin{figure}[!t]
	\centering
	\includegraphics[width=\linewidth]{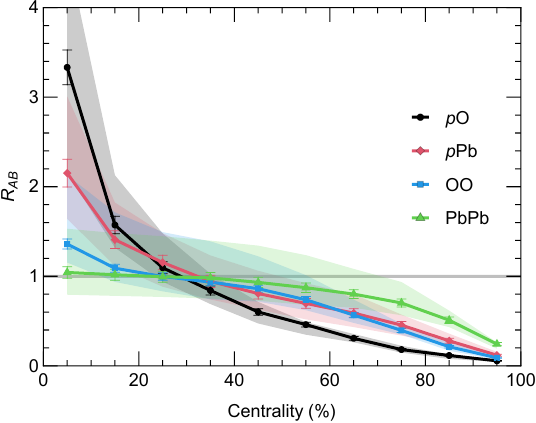}
	\caption{The unrescaled nuclear modification factor $R_{AB}$ of \cref{eqn:nuclear_modification_factor} for charged hadrons produced in a variety of collision systems as a function of centrality. IP-Glasma results are shown for \po collisions at $\sqrt{s_{NN}} = 9.62~\mathrm{TeV}$ (black circles), \ppb collisions at $\sqrt{s_{NN}} = 5.02~\mathrm{TeV}$ (red diamonds), \oo collisions at $\sqrt{s_{NN}} = 5.36~\mathrm{TeV}$ (blue squares), and \pbpb collisions at $\sqrt{s_{NN}} = 5.02~\mathrm{TeV}$ (green triangles). All $R_{AB}$ are computed with the value of $\Ncoll$ from IP-Glasma. Bars are statistical uncertainties, bands are systematic theoretical uncertainties due to sampling from the Bayesian posterior.}
	\label{fig:RAA_system_comparison_unrescaled}
\end{figure}

The centrality-dependent results of \cref{sec:centralitydependent_nuclear_modification_factor}
are plotted with the normalization offset of \cref{eqn:rab_normalization}
divided out sample by sample,
as in \cref{eqn:rescaled_raa}.
For completeness we show here the same results
without that rescaling,
i.e.\ the $R_{AB}$ of \cref{eqn:nuclear_modification_factor} as computed directly,
using the experimentally reported \Ncoll
together with the yields produced by each posterior sample.
\Cref{fig:RpA_vs_centrality_pPb_502TeV_unrescaled} shows $R_{p\pb}$,
\cref{fig:RAA_PbPb_cent_unrescaled} shows $R_{\pb\pb}$,
and \cref{fig:RAA_system_comparison_unrescaled} shows the comparison across all four collision systems,
to be compared respectively with
\cref{fig:RpA_vs_centrality_pPb_502TeV,fig:RAA_PbPb_cent,fig:RAA_system_comparison}.

In both \ppb and \pbpb the central values are essentially unchanged
and the systematic bands are wider by roughly a factor of four,
the same behavior found for the minimum-bias $R_{pA}$
in \cref{sec:nuclear_modification_factor},
except in the most peripheral class,
where the rescaling itself dominates the uncertainty (see \cref{sec:centralitydependent_nuclear_modification_factor}).
The band of \cref{fig:RpA_vs_centrality_pPb_502TeV_unrescaled}
covers the ALICE measurement in every centrality class except the most peripheral,
while that of \cref{fig:RAA_PbPb_cent_unrescaled}
spans $R_{AA} \simeq 0.8\text{--}1.5$ over $0\text{--}80\%$ centrality.

\section{The oxygen double ratio}
\label{sec:double_ratio}

\begin{figure}[!t]
	\centering
	\includegraphics[width=\linewidth]{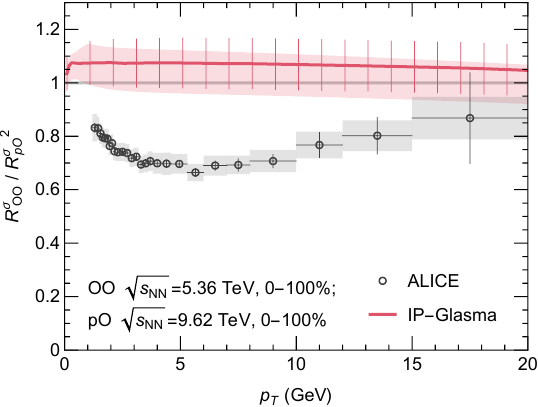}
	\caption{The double ratio $R^{\sigma}_{OO} / \left(R^{\sigma}_{pO}\right)^2$ of minimum-bias nuclear modification factors for charged hadrons, constructed from $0\text{--}100\%$ centrality \oo collisions at $\sqrt{s_{NN}} = 5.36~\mathrm{TeV}$ and $0\text{--}100\%$ centrality \po collisions at $\sqrt{s_{NN}} = 9.62~\mathrm{TeV}$, as a function of transverse momentum $p_T$. IP-Glasma results are shown in red and experimental data from ALICE~\cite{ALICE:2026zck} are shown in black. Red bars are statistical uncertainties, bands are systematic theoretical uncertainties due to sampling from the Bayesian posterior. Black bars are statistical uncertainties and shaded gray boxes are systematic uncertainties.}
	\label{fig:RAA_OO_over_pO_MB}
\end{figure}

The double ratio $R^{\sigma}_{AA} / {R^{\sigma}_{pA}}^2$ was introduced in Ref.~\cite{Jonas:2026yoz}
as a way to cancel the nPDF uncertainties
that dominate collinear-factorization predictions of $R_{AA}$,
and has since been measured in the oxygen systems by ALICE \cite{ALICE:2026zck}.
\Cref{fig:RAA_OO_over_pO_MB} plots $R^{\sigma}_{OO} / {R^{\sigma}_{pO}}^2$ for charged hadrons as a function of $p_T$.
We find
$R^{\sigma}_{OO} / {R^{\sigma}_{pO}}^2 \simeq \num{1.07(12:04)} ~\text{(sys.)} \pm 0.09 ~\text{(stat.)}$,
consistent with unity,
against a measurement that sits at ${\simeq}\,0.7$ over $5\text{--}10 ~\mathrm{GeV}$.

We present the double ratio here rather than in the main text
because the double ratio carries little information beyond \cref{fig:oxygen_MB} in our framework.
Our calculation does not use nPDFs,
so the uncertainty that the observable was constructed to cancel does not enter in the same form;
the normalization uncertainty that does affect $R_{AB}$
is already removed by the cross-section normalization of \cref{eqn:nuclear_modification_factor_sigma},
as shown in \cref{sec:inelastic_cross_section_parameters}.
Since we predict $R^{\sigma}_{pO} \simeq 1$,
the double ratio is in addition numerically close to $R^{\sigma}_{OO}$ itself.
We also see in \cref{fig:RAA_OO_over_pO_MB} that the double ratio does not reduce our theoretical uncertainty:
the four event samples entering the double ratio are statistically independent---%
the \pp references for \oo and \po sit at different $\sqrt{s}$---%
so the statistical uncertainty of the double ratio, $\pm 0.09$,
is larger than the statistical uncertainty $\pm 0.04$ of $R^{\sigma}_{OO}$ alone.
With a significantly higher-statistics campaign of IP-Glasma runs
we expect that the double ratio would remain consistent with unity.

\FloatBarrier

\bibliographystyle{apsrev4-2}
\bibliography{manual,ipglasma_hard_soft}

\end{document}